\documentclass{SNmult}

\usepackage{url}            
\usepackage{graphicx}       

\begin{document}
\title*{Interactive and Urgent HPC: State of the Research}

\author{Albert Reuther\orcidID{0000-0002-3168-3663} 
\and William Arndt\orcidID{0000-0001-8409-3133}
\and Johannes Blaschke\orcidID{0000-0002-6024-3990}
\and Christian Boehme\orcidID{0000-0002-4289-0465}
\and Nick Brown\orcidID{0000-0003-2925-7275}
\and Antony Chazapis\orcidID{0000-0002-4729-7396}
\and Bjoern Enders\orcidID{0000-0002-6009-6281}
\and Jens Henrik Goebbert\orcidID{0000-0002-3807-6137}
\and Robert Henschel\orcidID{0000-0003-2289-9398}
\and Julian Kunkel\orcidID{0000-0002-6915-1179}
\and Maxime Martinasso\orcidID{0000-0003-1849-1621}
\and Michael Ringenburg\orcidID{0000-0001-5119-0754}
\and Rollin Thomas\orcidID{0000-0002-2834-4257}
}

\institute{Albert Reuther \at MIT Lincoln Laboratory, \email{reuther@ll.mit.edu}
\and William Arndt \at Lawrence Berkeley Lab, \email{warndt@lbl.gov}
\and Johannes Blaschke \at EIT Oxford, \email{johannes.blaschke@eit.org}
\and Christian Boehme \at GWDG, \email{christian.boehme@gwdg.de}
\and Nick Brown \at EPCC, \email{n.brown@epcc.ed.ac.uk}
\and Antony Chazapis \at FORTH, \email{chazapis@ics.forth.gr}
\and Bjoern Enders \at Lawrence Berkeley Lab, \email{benders@lbl.gov}
\and Jens Henrik Goebbert \at FZ-Juelich, \email{j.goebbert@fz-juelich.de}
\and Robert Henschel \at Indiana University, \email{henschel@iu.edu}
\and Julian Kunkel \at GWDG and Universität Göttingen, \email{julian.kunkel@gwdg.de}
\and Maxime Martinasso \at Swiss National Supercomputing Center, ETH Zurich, \email{maxime.martinasso@cscs.ch}
\and Michael Ringenburg \at Allen Institute, \email{mike.ringenburg@alleninstitute.org}
\and Rollin Thomas \at Lawrence Berkeley Lab, \email{rcthomas@lbl.gov}
}

\maketitle

\abstract*{
When we think of how we use smartphones, e-commerce, collaboration platforms, LLMs, etc., most of our interactions with computers are interactive and often urgent. Similar trends of interactivity and urgency are coming to HPC, with applications from simulations to data analysis and machine learning requiring more parallel computational capability and more interactivity. This chapter overviews the progress made so far along with some vectors of what the path forward will bring for greater integration of interactive and urgent HPC policies, techniques, and technologies into our HPC ecosystems.
}

\abstract{When we think of how we use smartphones, e-commerce, collaboration platforms, LLMs, etc., most of our interactions with computers are interactive and often urgent. Similar trends of interactivity and urgency are coming to HPC, with applications from simulations to data analysis and machine learning requiring more parallel computational capability and more interactivity. This chapter overviews the progress made so far along with some vectors of what the path forward will bring for greater integration of interactive and urgent HPC policies, techniques, and technologies into our HPC ecosystems.}

\section{Introduction}
\label{sec:introduction}

Computing has become ubiquitous in our everyday lives. In our pockets, handbags, and backpacks, most of us carry greater computational capabilities than supercomputers from just a few decades ago. 
Over the past decade or so, we have become very familiar with, and rely on, the completely interactive user experience with our smartphones, tablets, and laptops to answer any question that entered our mind moments ago. 
Ultimately computing is fundamentally interactive; it is how almost all of the world interacts with their computers. 
A similar transformation toward interactivity is occurring in highly parallel research and scientific computing, commonly termed high performance computing (HPC)/supercomputing (usually used interchangeably), in which tens, hundreds or thousands of high-end servers are used in parallel to solve massive simulation, data analysis and machine learning training tasks. 
However, HPC centers still have a way to go in their transition to fully embracing interactivity. 

Interactive computing first emerged a few years after the dawn of modern computing during World War II. The first demonstration of an interactive computer session was conducted in 1963 on the second transistor experiment system, TX-2, at MIT Lincoln Laboratory based on the PhD thesis work of Ivan Sutherland (subsequent co-founder of Sun Microsystems). In his Sketchpad demonstration, he showed several capabilities that we now take for granted, such as drawing geometric figures on the screen that affected the computations of the computer\footnote{\url{https://www.youtube.com/watch?v=6orsmFndx_o}}. 

The first supercomputers started emerging in the 1980s, and have always been considered very high value assets. Similar to expensive experimental equipment like telescopes, they require proposals for allocations, planning, and meticulous scheduling to execute jobs against management-granted compute time allocations. Furthermore, the types of simulations that were, and still are, executed on these supercomputers have required extensive analysis and planning to determine the parameters, initial conditions, and simulation meshes with which to execute each parallel job. 
Because the execution time of these parallel jobs often spans hours, days and even months, there is usually little churn in the scheduler allocation of resources to executing jobs, thereby allowing extensive scheduler analysis and prioritization of which jobs to run next from the jobs waiting in the queue. 
This type of execution is commonly referred to as batch scheduling, and modern supercomputers are built around this design: jobs are prepared on a login node, submitted to a batch queue, and then executed on compute nodes with no direct user access (i.e., offline). 

In the past two decades, however, it has become apparent that a significant subset of HPC jobs are far more effective when they are run online, rather than offline, with respect to the user. These jobs are typically known as interactive, and their common denominator is being sensitive in regard to their start and/or completion times as well as potentially benefiting from some steering from the user whilst they are running. 

Interactive HPC involves users \emph{being in the loop} during job execution where a human is monitoring a job, steering the experiment, or visualizing results to make immediate decisions about the results to influence the current or subsequent interactive jobs~\cite{reuther2005technology}. 
Interactivity is often the first step in scaling to larger models or datasets, as well as being part of an agile development and testing cycle. 
This can include, for example, intuition-forming exploratory parameter scans, \emph{what-if} scenarios, and investigations of extreme cases. These are all at the core of the scientific discovery process, and furthermore, the preparation and debugging of state-of-the-art simulations has become increasingly difficult as data sizes have grown when exclusively using the traditional batch queue approach. The conventional approach is further complicated by the fact that modern data sizes and simulation complexity means that even the most powerful on-site user workstations are no longer capable of supporting interactivity during simulation preparation or repeated data transfers between the scientist’s laboratory and remote supercomputing sites. Moreover, numerous computationally-demanding emerging scenarios could all benefit from supercomputing resources, but they are often hard to realize in the traditional, batch-oriented workflow. 
Further, it is not only computational workloads, but also data analytics and machine learning workflows that frequently require interactive exploration of large data sets. 

Urgent HPC is an emerging supercomputing use-case and this involves immediate data or actions that will fail if the job is not run within strict time constraints. Consequently, urgent HPC is similar to deadline scheduling where an HPC job has to deliver results by a given time. However, this is complicated by the potentially highly dynamic and unpredictable nature of this workload, depending on whether the event to which the job is couple can be planned or not. 
The extensive use of supercomputing during the global COVID-19 pandemic~\cite{hardy2021lessons,cheng2021real} and recent bouts of extreme climate events~\cite{brown2019role,lovholt2019urgent,goubier2020fast} are examples of this rapidly growing area of HPC usage with a significant societal benefit in tackling emergency scenarios. Such activities have demonstrated the need to make urgent and accurate decisions for complex problems, and combining interactive computational modeling with near-real-time detection of unfolding disasters results in a powerful tool that can help emergency responders make life-critical decisions for disaster response~\cite{mandel2019interactive}.  In a similar vein, supercomputers can be used in the operating room guiding the surgery process, provided urgency requirements are met~\cite{Zhou2020}.  Ultimately, the objective here is to exploit HPC to deliver significant societal benefits by saving lives and reducing economic loss. 

In recent years numerous efforts to further the state of the art in interactive and urgent use of HPC resources have been undertaken.  
Indeed, much progress has been made in bringing interactive and urgent HPC capabilities online at a number of supercomputing centers across the world. 
However, the HPC community has been slow to adopt interactivity across the board in a widespread and consistent fashion. 
In this chapter we take a snapshot inventory of the current state of the practice in enabling and implementing interactive and urgent HPC capabilities across the world with a particular lens of a series of birds-of-a-feather sessions and workshops that the author team has been organizing over the past decade. From that snapshot inventory, we proceed in casting a path forward before drawing the chapter to a close with a conclusion. 

\section{The State of the Practice}
\label{sec:currentstate}

Over the past several years, numerous birds-of-a-feather meetings and workshops have been hosted at IEEE/ACM Supercomputing and ISC-High Performance conferences, which have focused on the unique challenges of enabling interactive use and urgent use of HPC systems. These events have enabled the authors to connect with a wide range of supercomputing stakeholders. The results of these discussions are captured in this chapter to synopsize the current state of practice in regard to interactive and urgent HPC capabilities. It should be highlighted that the selection of research included in this chapter is made on the basis of these discussions and the authors' expert opinion.
The current state is divided into the following topics: \emph{organizational and system policy}, \emph{scheduling techniques}, \emph{system infrastructure and tools}, \emph{data management}, \emph{performance benchmarking}, \emph{user support} and \emph{user case studies}.

\subsection{Organizational and System Policy}

When a supercomputing center considers introducing or expanding the number of compute nodes that are available for executing interactive and urgent sessions, they are usually faced with several organizational and management policy concerns. 
These policy concerns are rooted in the traditional view that purchasing and maintaining an HPC system is very expensive and that jobs on such a machine must be scheduled to achieve high utilization. (Over 95\% utilization is common.) Ultimately this is then used to justify the supercomputer's purchase and recurring operations costs to the funders. In order to achieve close to full utilization, jobs are scheduled in ``batch mode'', where users submit their jobs into a queue of pending jobs which then wait until resources are available to execute each one. Further, having a queue of possible next jobs to run enables the scheduler to more optimally allocate jobs to computing resources. 
Wait times can range from minutes to hours to days to weeks and this depends on many factors. Consequently, batch scheduling is the antithesis to interactivity, because these two modes represent two entirely different approaches to operation. By contrast, interactive and urgent workloads require a collection of idle nodes which must be available to quickly launch such jobs, and this leads to a lower utilization percentage overall.  
Some centers have introduced a wider selection of metrics based upon return on investment (ROI), replacing the utilization metric that is more focused on user productivity~\cite{reuther2006high}. For example, the DARPA HPCS program in the early 2000s, from which the aforementioned ~\cite{reuther2006high} paper came, and a BoF session at SC-13 on cost-benefit analysis of HPC attempted to address this policy concern. 


In recent years, many organizations have de-emphasized the importance of keeping all computing resources utilized all of the time. This has come in different forms. 
Some centers have changed the goal value for utilization to a lower percentage than 90\% or 95\%. Others have made lists of caveats; e.g., idle times that do not count against utilization, job types that are exempt, or compute nodes that are not included in the utilization metrics. 
Even leadership class HPC centers have taken advantage of scheduling policies that enable interactive and urgent workloads. For instance, at the workshop at ISC-2019, Jack Wells explained how ORNL employed scheduler policies and queue management to enable urgent jobs on the Summit supercomputer~\cite{wells2019scheduling}. 

Also in recent years, many organizations have seen a growing need for providing HPC
resources that are responsive enough to enable urgent and interactive computing.
Often such resources are known by the term ``on-demand'' or ``real-time'' computing. A
careful balance must be struck here as batch computing is incredibly successful
at maintaining high utilization, which in turn enables more research to be
performed on valuable machines. Conversely, providing some on-demand computing resources for certain urgent and interactive research projects has also proven to be valuable by yielding high-impact research. For example, it has become
unthinkable that the Jupyter research notebook platform is  not available on
modern HPC systems, and Jupyter notebooks are often only allowed to run on such on-demand partition of compute nodes. Other applications in this category include research desktops, HPC web portals, and more, which will be discussed further in Section~\ref{subsec:technology}. 
While it might be counterintuitive to system operators, a fairly consistent
observation among HPC practitioners that has been discussed throughout the community is that one needs only ``sacrifice'' a small
percentage (less than 10\%) of total compute nodes in order to successfully accommodate most
interactive users. However, along with those conversations come discussions about whether 10\% is really enough and what should be done when 10\% is not enough. On the other end of the spectrum, some centers have embraced interactive and urgent HPC by making an organizational policy decision to prioritize interactive jobs over batch jobs on the majority of their HPC assets at organizations including MIT Lincoln Laboratory~\cite{mullen2018lessons}, the U.S. Department of Energy Superfacility~\cite{enders2020cross,Superfacility}, and the American Science Cloud (AmSC)~\cite{wilkinson2025designing}. 
As described in this section, there is a tradeoff between utilization and effectiveness for interactivity, with a middle ground likely being the most effective approach for most HPC centers and their users.

\subsection{Scheduling Interactive and Urgent Jobs}
\label{sec:scheduling}

How jobs are actually scheduled on HPC systems plays a large role in their suitability for interactive and urgent workloads. All HPC systems have at least one shared login node upon which users can execute some interactive work if required. However, given that login node(s) are shared between all users and crucial for developing, testing, preparing and submitting jobs, there is a risk of contention where computationally intensive interactive workloads can quickly overwhelm this resource. Consequently, larger systems tend to introduce a debug queue which provides interactive debugging, but usually the resources allocated to that queue tend to be fairly limited and sessions are usually short (30 minutes to two hours maximum) depending on the policy of the supercomputing center. Neither of these two approaches adequately accommodate full interactive and urgent jobs of any complexity, scale and duration.  
Some centers rely on reservations to provide resources that can accommodate interactive and urgent sessions where the debug queues are not suitable; however, the limitation here is that these jobs must be known a priori, which does not always work for unpredictable and dynamic workloads especially in urgent stream computing nor in high-priority/emergency user-steered interactive code development. Other centers go further by identifying certain jobs that can be preempted, and these share a queue with interactive and urgent jobs, typically with owners of such preemptable jobs given preferential treatment such as prioritised scheduling or reduced cost~\cite{minami2025physical,byun2020best}. More centers experimented with temporally-changing dynamic resource allocations~\cite{sala2025novel} and short job execution windows with restarts~\cite{mcglothlin2025implementing}. Yet other centers have experimented with various policies for scheduling urgent jobs among batch jobs~\cite{klein2020interactive,maheshwari2025evaluating}, and how to schedule urgent jobs consistently among prototype and production HPC systems~\cite{etz2025enabling}. 


Much of the published work on scheduling interactive and urgent jobs on HPC clusters has focused on the currently prominent schedulers, Slurm and Kubernetes, with early efforts also leveraging HTCondor, IBM Platform LSF, GridEngine, PBS, and Torque. For the most part, Slurm has become the defacto standard HPC scheduler, while Kubernetes has become the standard for cloud and containers. 

Researchers have evaluated various feature sets for launching interactive jobs with regard to the types of jobs that these schedulers support. For instance, high throughput computing and interactive jobs compared against large scale, optimally mapped jobs~\cite{reuther2018scalable}. 
It has been consistently observed that there are tensions between how quickly large jobs are scheduled versus smaller/interactive jobs. Further tension exists between multi-node synchronously parallel jobs and block synchronous (high throughput) jobs. A good balance between all of these is difficult or perhaps impossible to find, especially as job characteristics change over months and years. Furthermore, the behavior of the scheduler often depends on the policy imposed by the specific supercomputing centre, meaning that the quality of service for specific jobs types can change substantially when moving from one HPC system to another.

This tension has driven the development of strategies and implementations that make interactive and real-time software appear like batch jobs, but operate interactively. For instance, BatchSpawner enables Jupyter notebook servers to run in a batch job, and KubeSpawner allows them to run as a pod under Kubernetes on demand, the latter arguably being a more natural approach which has seen greater adoption. Pilot jobs~\cite{turilli2018comprehensive} that load tasks from an external database and process these create the illusion of queue-less real-time analysis as a service, but handling this within the queue system itself would be a better approach. 

Another approach has been implemented in Kubernetes, where a new scheduler is created and sends jobs to optimized sub-schedulers for different types of jobs depending on a variety of characteristics such as job size. Different queues and priorities are exploited to effectively schedule these jobs, and a similar feature is being developed for the Flux scheduler~\cite{ahn2020flux}, which aims to succeed Slurm. 

Urgent jobs with latency requirements in the single digit second range, such as LLM inference in chatbots, cannot usually be accommodated by single job scheduling. Typically, the scheduling task required here is demand based scaling of a permanently available service. This type of scheduling is provided by Kubernetes autoscaling, and to integrate this service scheduling with classical HPC job scheduling on the same system, different approaches have been developed. For example, Kubernetes has been extended by the batch scheduler Volcano \cite{volcano}, and there is ongoing work on provisioning Slurm clusters under Kubernetes using the MPI Operator \cite{mpi-operator}, Soperator \cite{soperator}, or SUNK {\cite{sunk}} (SlUrm oN Kubernetes).
Slurm-operator \cite{slurm-operator} (part of SchedMD's Slinky projects \cite{slinky}) can manage both a containerized Slurm deployment that operates totally inside Kubernetes, or a hybrid setup where some components (\emph{i.e.}, compute workers) run on bare-metal.

\begin{figure}
    \centering
    \includegraphics[width=0.5\linewidth]{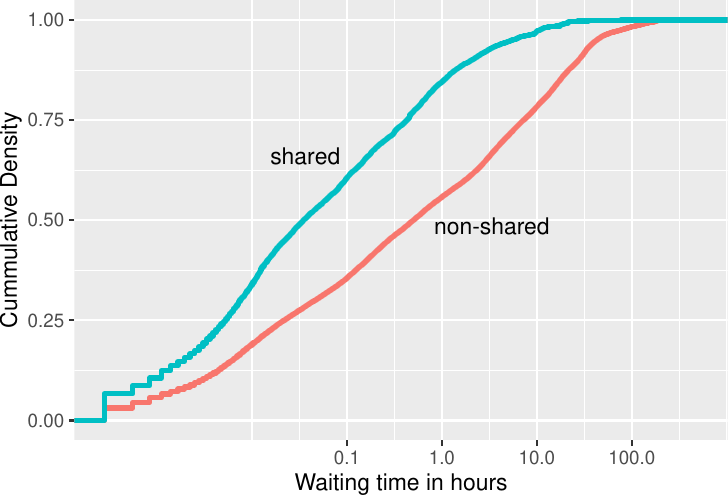}
    \caption{Waiting time for jobs executing on shared nodes vs. exclusive}
    \label{fid:shared}
\end{figure}

\begin{figure}
    \centering
    \includegraphics[width=0.5\linewidth]{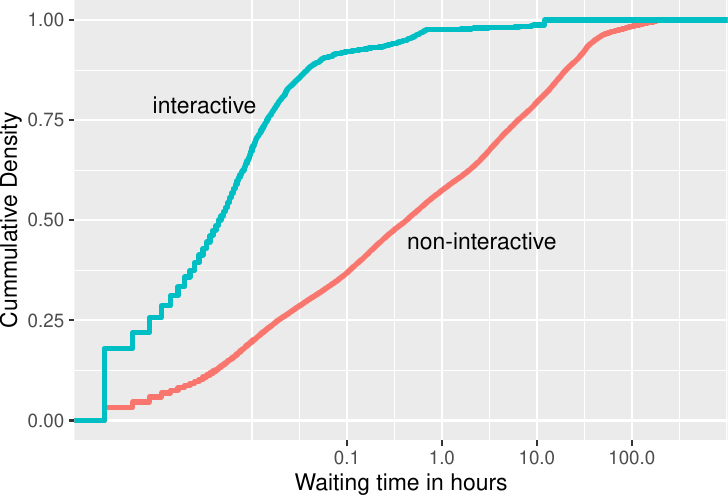}
    \caption{Waiting time for interactive vs. non-interactive jobs}
    \label{fid:interactive}
\end{figure}

With a number of interactive and urgent HPC scheduling strategies to choose from, the Gesellschaft für Wissenschaftliche Datenverarbeitung mbH Göttingen (GWDG) has developed metrics and is collecting and analyzing data toward better comparisons of the strategies~\cite{kunkel2024challenging}. 
Data from GWDG job scheduling is provided in Figures~\ref{fid:shared} and ~\ref{fid:interactive}. During a one-year period, about 3.5 million jobs were executed on the various clusters at GWDG. 
A handful of jobs that were waiting longer than 500 hours were removed from the data as these were special circumstances.
We provide shared access or exclusive access on nodes as options --- about 220k jobs were run on shared nodes.
Similarly, jobs can be requested on interactive partitions, which use both node sharing and node oversubscription - about 16k jobs were interactive or Jupyter jobs. 
We can see that jobs requested in an interactive queue are dispatched significantly faster, with a similar benefit on shared nodes.
While GWDG started to look into preemptible jobs, it is not yet used much. Currently, we investigate options to incentivize it.

Other research has been undertaken that explores the use of Machine Learning (ML) to better understand and help predict batch queue wait times~\cite{brown2024predicting,whitton2025we}. While this shows promise, there are a series of challenges that can be difficult to overcome, and one of the most significant is that, as users tend to specify the maximum rather than absolute runtime of their job in submission scripts, it can be difficult to obtain an accurate estimation of the amount of work in the queue. Nevertheless, this approach was used by~\cite{flatken2023vestec} in designing their urgent computing system which federated across HPC systems in a range of geographical locations. Using ML to estimate the job start time for each machine, mapping of jobs to supercomputers was based upon a combination of this estimated start time along with machine suitability and the overhead involved by any required movement of data.






\subsection{Technology for Interactive and Urgent Capabilities}
\label{subsec:technology}

Effective scheduling and execution policies are fundamental for facilitating interactive and urgent computing, yet they are insufficient in isolation. Equally crucial are environments and tools that have been specifically designed to support interactive and urgent workflows. This requirement is addressed through interactive and programmatic environments tailored for workflow construction. Users engaged in interactive and urgent computing tasks typically require environments that enhance their productivity; these include code development, debugging, and data analysis along with integration across one or more HPC systems. Such environments can be categorized into integrated application environments, web portals, and research desktops;  APIs that enable the development, customization and automation of complex computational workflows; tools to work across heterogeneous environments; and streaming workflow tools.

\subsubsection{Interactive and On-Demand HPC Environments}

Initial efforts to build interactive and urgent user environments involved the creation of custom solutions by HPC providers and researchers. One avenue of these initial efforts was in enabling parallel MATLAB~\cite{choy2005parallel} on HPC systems such that the user's interactive MATLAB sessions mirrored the responsiveness and richness of desktop MATLAB sessions~\cite{reuther2005technology}. A notable instance of such an environment is the configuration of a MATLAB framework on a computing cluster, utilizing MatlabMPI, pMatlab, and gridMatlab libraries for parallel processing~\cite{reuther2004llgrid}. These MATLAB processes were initiated via the cluster's job scheduler~\cite{reuther2007technical}. Concurrently, analogous systems such as Star-P~\cite{choy2002matlab}, Parallel MATLAB Toolbox~\cite{gradfreilich2023interactive}, gridMathematica~\cite{maeder2006gridmathematica}, and Techila~\cite{lin2018practitioner} were developed and broadly accepted, each aiming to provide users with the tools necessary for interactive computation and data processing in an HPC context.

A similar avenue that other HPC centers took was integrating and enabling Jupyter Notebooks for interactive and urgent code development, debugging, data exploration, etc. in HPC centers. These endeavors included integrating Jupyter tools (i.e., Jupyter Notebooks, JupyterLab, JupyterHub) into the software stack, scheduler, and user environments~\cite{farrell2018deep,goebbert2018enabling,henderson2020accelerating,ragan2024enabling,werner2024jumper}. 
More recently, feature-rich Integrated Development Environments (IDEs) such as Visual Studio Code~\cite{raess2023teaching,zhukov2025onboarding} are being integrated into HPC environment offerings. These environments, while quite powerful, require specific expertise and resources to configure for particular applications. 

As these interactive and urgent user environments gained traction, several HPC centers independently developed portals to simplify access to such tools. Early examples include the portals created by the Department of Defense High Performance Computing Modernization Program (DoD HPCMP)~\cite{atwood2016secure}, MIT Lincoln Laboratory~\cite{prout2017mit}, and the San Diego Supercomputer Center (SDSC)~\cite{sakai2021experiences}, which typically utilize proxy-based mechanisms to provide desktop users with browser-based access to integrated tools such as JupyterLab. 
Building on these initiatives, researchers at the Ohio Supercomputing Center and the University of Virginia introduced Open OnDemand~\cite{OpenOnDemand}. This open source framework furnishes HPC centers with the underlying structure and portal capabilities required to offer a diverse array of interactive and urgent HPC services via a web interface. Open OnDemand serves as a pivotal development in democratizing access to HPC resources, streamlining the user experience~\cite{settlage2019open,settlage2020tools,sadek2023open}. An extensive survey of HPC web portals including feature comparisons is captured in~\cite{calegari2019web}. 




Alternative approaches have facilitated the integration of research desktops, which are environments provisioned directly on HPC systems that provide seamless access to HPC services. The research desktop paradigm allows users of varying expertise to leverage computational power and storage capacity with the familiar ease of a traditional desktop interface.
These research desktop servers function similarly to the login nodes of an HPC system where they are situated in close proximity to HPC clusters, have cluster file systems mounted, and act as submission hosts for batch processing~\cite{henschel2024use,lindeman2024interactive}. Unlike login nodes, which typically offer only SSH access, research desktop nodes provide remote desktop access using solutions such as ThinLinc, NoMachine, X2Go, or FastX. In addition, they are configured to allow the execution of sustained graphical applications, such as MATLAB, RStudio, or Visual Studio Code, thus accommodating a range of computational tasks within a user-friendly environment.

Research desktops can also be integrated with on-demand HPC environments. For instance, the GWDG HPC center uses a JupyterHub instance to offer Jupyter notebooks, RStudio, a web-based Visual Studio Code environment with AI tool integration, and multiple Linux desktops flavors as virtual environments, all of which are supported by an HPC backend. Tasks from the different interfaces are started as Slurm jobs on dedicated resources and are packaged as Apptainer containers. If selected, the jobs can use GPU resources to provide 3D acceleration inside their virtual environments for demanding visualization tasks via VirtualGL and VNC.

\subsubsection{Web-based Application Programming Interfaces}

In parallel to the development of interactive and On-Demand HPC environments,
some organizations have pioneered the development of web-based programmatic
interfaces to HPC resources, empowering the scientific community to construct
bespoke portals and workflow engines. These HPC centers offer RESTful API access
to their services, principally exposing data transfer and job submission
functionalities. Utilizing these APIs, along with their language-specific
bindings, researchers can seamlessly integrate interactive and urgent HPC
workflows, customizing them to optimize productivity.
An example is the Materials Cloud~\cite{MaterialClouds} platform, which facilitates interactive studies of material properties through a dedicated web portal. It harnesses the AiiDA workflow engine~\cite{AiiDA} 
written in Python and the Python bindings of FirecREST~\cite{firecrest}, 
a web-facing APIs. Such APIs simplify support requirements for HPC centers as they can provide a single interface to cater to diverse scientific requirements.
Additionally, various other APIs such as HEAppE~\cite{HEAppE}, the Superfacility API~\cite{Superfacility}, and HPCSerA~\cite{koehler2022}
have been developed. Some APIs also provide serverless computing endpoints, which allow for service-based calls, further extending the functionality and flexibility of HPC resources for user-driven computational tasks.

\subsubsection{Enabling HPC Across Heterogeneous Environments}

Another related ongoing theme is the convergence of HPC and Cloud computing. From the HPC perspective, there are a breadth of tools and services already available for the Cloud in HPC such as KNoC~\cite{maliaroudakis2022interactive}, which manages container lifecycles on remote HPC systems, and HPK \cite{hpk}, which enables users to deploy Kubernetes within an HPC environment. Using a custom kubelet, this translates the container's lifecycle to Slurm and Apptainer commands. Moreover, concepts are introduced into the HPC world to provide Cloud flexibility on top of HPC systems. An example of this is the model of versatile software-defined cluster~\cite{vcluster}. By leveraging network segregation and containerization, this model offers a more flexible and adaptable service delivery to HPC users, aligning with Cloud computing paradigms while maintaining the core strengths of vertically integrated software stack of HPC systems.


\subsubsection{Streaming Workflows}

In response to the ``data tsunami'' from modern scientific instruments \cite{rao2020deluge, Spurgeon2021-ym}, brought
about by advances in both detector design and faster measurement rates,
real-time data analysis workflows designers are increasingly looking at
streaming raw data directly from detector to HPC node (or even GPU) memory. 
While the community at experimental user facilities, such as light sources, has long since been moving away from portable storage devices to networks and data servers, many setups have not been able to keep pace with processing the data at a rate comparable to what the instruments produce. In order to keep up with processing of the generated data, partnerships have sprung up where HPC facilities take over data processing to relieve computing pressure at the edge~\cite{Superfacility}. This has, however, not improved upon the timeliness of such processing. In fact, it made it worse. Having to wait in the scheduler's queues for your job to run or waiting for the data to be transmitted adds to the latency even with the compute power of a HPC facility at your disposal. While the facility can try to accelerate the processing with priority scheduling or reservations, the workflow is hampered by piping its data through shared storage systems where a few bad actors can lead to a severe and unexpected degradation of performance (both latency and bandwidth) in addition to the latency introduced by chain of function calls from (remote) file I/O.

In \cite{swelborn2024mam} the authors report how they could stream data
directly between instrument and HPC node memory using {\it ZeroMQ}. Streaming data directly into compute nodes eliminates entirely the performance risks associated with any shared component in the data's way. An exception is the network but the bottleneck here most often is on the WAN side and would primarily affect throughput. While opting for streaming instead of file transfer is very effective in addressing latency, it comes with its own caveats. First, in absence of any buffering mechanism, it requires the whole workflow to be synchronous, i.e. all components (compute, network, instrument) need to be up and functional for the workflow to execute. In fact, streaming would make the processing at the HPC facility an integral and essential part of the scientific workflow. Second, streaming often requires a complete retooling of the pipeline. In order to unlock the benefits of a streaming workflow, file I/O should be avoided along the entire data path. The could mean tapping into the data stream at the source, e.g., at a detector. And the vendor of that detector must be open to provide the interfaces for that. Third, a direct connection between the compute node and the data source might raise security concerns on either side of the pipeline. This requires that a trusted broker service can be established between the two sites for the duration of the workflow.

While these challenges are real, they are not insurmountable and the payoff may be huge. In \cite{welborn2024high-abe}, the authors report a performance boost of up to 5x to 14x, pushing the workflow into the realm of real-time feedback for the experiments at the user facility. It means that the user was presented almost immediately with the processed results, allowing this feedback to influence their next decision.

\subsection{Data Management}
\label{sec:CurrDataManagement}

As was highlighted in Section~\ref{sec:introduction}, the need for urgent and interactive resources arises from the analysis of data. Therefore effective data management can be seen as a lynchpin of any urgent and interactive workflow, and HPC facilities are aware of this having started to develop infrastructure to better manage high volumes and rates of data. In this section we highlight the areas which we have identified as being most impactful to urgent and interactive HPC workflows.

Today, most HPC facilities deliver data storage and management through the deployment of large shared file systems along with a mechanism that maps their user landscape to Unix users and file groups. 
Additionally, object storage interfaces such as S3 are provided as cold storage or for data sharing.
The engineering trade-offs that are made to enable large-scale shared storage can lead to unpredictable performance which, in turn, can cause workflow stalls due to up to 100 times slower performance then expected. 
While many teams' workflow jobs will be impacted by slow I/O, users with interactive or urgent workflows are impacted  particularly severely as important deadlines would be jeopardized \cite{blaschke2021realtime}. 
These users require reliable access to compute resources; this also includes all I/O associated with the interactive/urgent workflow. 

This concern on the quality of service for data I/O can readily be expanded to all networks that are touched in the path of the data flow. For many urgent use cases, where data is generated externally but analyzed locally, the data must be communicated to the HPC facility in time for the compute job to run. Indeed, waiting for data to arrive can result in unacceptable delays to job initiation and is akin to waiting in the scheduler queue for nodes.
The temporally unreliable nature of shared file systems, and the asynchronicity associated with file transfer to and from the HPC storage infrastructure, has forced some teams that prioritize fast feedback to explore memory-to-memory streaming solutions. This is where shared filesystems are avoided until the end of the pipeline and whilst promising these solutions are novel from an HPC point of view, they might require exceptions to firewall rules and thus face resistance from a security and policy perspective at HPC facilities. Furthermore, as some HPC systems are designed with no direct path between compute nodes and external networks, such designs will not only introduce additional latency, but also additional potential points of failure.

Another issue is the coordination of data transfer with computation. Whilst some workflow management tools have begun to develop features and best practices, the current landscape is dominated by each application implementing its own solution with little guidance from HPC centers. These solutions normally take the form of regularly checking the state of the file system, or polling an API as to the state of a given transfer. This almost always happens independently of the HPC center batch scheduler, and such a lack of coordination results in additional delays in starting data analysis and leaves tracking data provenance entirely to the workflow tool.

Finally, interactive and urgent workflows face challenges with HPC data management around more of a social or trust nature across scientific domains. For instance, data at experimental facilities are often generated by a machine rather than a person and may be attributed to a group of people that are without a corresponding file group at the HPC facility or that do not even have user accounts. Moreover, the team could be of a transient nature and be dissolved after the data has been analyzed and published. Intense competition in the sciences also forces the embargoing of data until it is ready for publication. HPC facilities today are not well equipped to serve this user group as projects and group mappings may be made through a separate process where the Principal Investigator (PI) or group for the HPC accounts differs from the PI or group that "owns" the data. Strategies to map social groups to HPC groups or users vary, and one group might opt to have everything done under a single "machine collaborative" account to which there exists a separate access mapping for a few select administrator accounts. Another strategy is to go all-in where every external user applies for an HPC account. Both approaches are not ideal, as the former raises questions about acceptable use and traceability of user actions while the latter greatly inflates the number of users on an HPC system and puts the management burden onto HPC staff. Indeed, in ~\cite{gibb2020bespoke} the authors captured data from a variety of external sources, and this was then owned by a single processing user on the HPC system where it was processed. However, because multiple people could use the upstream system to drive the HPC, there was a potential for account sharing, therefore this approach required explicit approval from the HPC center due to their general user policy.

\subsection{Performance and Benchmarking of Urgent and Interactive HPC}

As implied by the name itself, performance is a critical component of High \emph{Performance} Computing (HPC).  Given this, the HPC community spends a great deal of time developing, running, and analyzing performance benchmarks such as LINPACK~\cite{linpack}, HPCG~\cite{hpcg}, Graph500~\cite{graph500}, Green500~\cite{green500}, and HPCChallenge~\cite{luszczek2005introduction}.  Most of these HPC benchmarks focus on application throughput, and for traditional batch-based HPC jobs this metric makes sense as it captures what is arguably the most impactful aspect of batch job performance.  However, for urgent and interactive HPC use cases, factors other than throughput must also be considered.  In particular, in urgent scenarios, the time to first response is often key, which touches on aspects such as latency and scheduling preemptibility. In interactive use cases, user response latency is critical. The MIT Lincoln Laboratory Supercomputing Center team has been keen to keep interactive and urgent job launches to only a few seconds. Their early work on pMatlab and gridMatlab required that jobs launch in less than five seconds to maintain the same interactivity and responsiveness as desktop MATLAB sessions~\cite{reuther2007technical}. As the team added features and capabilities, they kept benchmarking to measure the launch time latency for those features and capabilities, including dynamic KVM/QEMU virtual machines~\cite{jones2016scalability}, HPC web portal services~\cite{prout2017mit}, machine learning frameworks~\cite{reuther2018interactive}, Windows emulation (WINE) environments~\cite{jones2018interactive}, and full-node task scheduling~\cite{byun2021node}. 

However, outside of limited examples such as those mentioned above, the HPC community has not traditionally focused on alternative metrics like latency and time to first response. As such, benchmarking and performance analysis of Urgent and Interactive HPC is a potentially fruitful "green field" research area.  Fortunately, there is relevant work from other communities that can be leveraged and built on.  For example, researchers in the operating systems community have developed benchmarks of the latency of kernel schedulers for application to real-time systems~\cite{abaffy2009,fan2020,wong2008,interbench}.  Similarly, frameworks such as VNCplay~\cite{vncplay} have been developed to measure response times for GUI-based applications. Frameworks such as Eithne~\cite{eithne} have also been developed to benchmark edge devices that are often used to collect data to feed Urgent HPC use cases. In addition, inference latency is extensively studied in the context of AI models, in particular in chatbot and question-answering style use cases~\cite{mlperfinference,mlperfclient} where Time To First Token (TTFT) is one of the most important metrics.  In this section, we survey some of this work and suggest directions for future work in benchmarking and performance analysis for urgent and interactive HPC.

The Interbench~\cite{interbench} benchmark is widely used to measure the interactive performance of various applications under various system loads.  It contains a variety of emulated use cases (audio, video, gaming, and a customizable use case), as well as a variety of simulated background loads such as memory pressure, compilation, and heavy I/O.  The operating systems and real-time systems communities have used this to evaluate the impact on interactivity of various scheduling algorithms and OS kernels~\cite{abaffy2009,fan2020,wong2008}.  While Interbench is primarily focused on measuring localized effects and interactivity on a single desktop or server, there are parallels to measuring the performance of interactive and real-time/urgent jobs on HPC systems. Similarly to OS schedulers on a local system, HPC resource managers schedule jobs across cluster resources in an HPC system.  While the time scale granularity is significantly larger, similar types of policies and considerations apply.  Many types of applications with varying constraints run on HPC systems, and they may operate under differing loads depending on the system, time, and seasonal impacts (e.g., conference deadlines).  Given these parallels, a similar benchmark to Interbench, but designed for HPC clusters, with HPC workload types and typical HPC background load patterns, would be a valuable contribution to the interactive and urgent HPC community.

Urgent, real-time, and interactive jobs often involve human-in-the-loop interactions with a graphical user interface (GUI).  These types of interactions have traditionally been challenging to model with standard benchmarking tools.  The VNCplay framework~\cite{vncplay} was built to address these gaps in performance testing.  VNCplay records user sessions and replays them under a variety of system configurations and load settings, and evaluates performance by comparing times between designated screen updates in the replayed sessions. This tool could also be used to measure response times and interactivity in urgent or interactive HPC jobs with GUI-based front-ends by running locally on the system where the front-end is executing.

In addition, urgent HPC use cases often involve working with data generated by edge devices, such as weather data, heatspots for wildfires, and seismology data. In these cases, preliminary processing typically occurs on small, power-constrained compute resources located near the edge.  Since these devices supply key input data to urgent HPC workloads, understanding their performance is critical to understanding the performance of the full workload.  To this end, the Eithne framework~\cite{eithne} was developed to benchmark low-power, highly parallel micro-core architectures which are often used in edge applications. The authors originally focused on LINPACK for floating-point performance and DFT and FFT benchmarks for relevance to disaster scenarios, but the framework is extensible and other benchmarks can be added. An example of using Eithne for urgent computing edge devices is from an EPCC team who benchmarked several microcore CPUs for detecting disaster signals at the edge~\cite{jamieson2020benchmarking}. In addition, a team from the University of Utah compared the performance of cloud-based and edge-based processors to determine where it would be most effective to process certain types of data~\cite{balouek2021evaluating}. 

AI model inference is often a key component of urgent and interactive HPC workflows.  For example, AI weather models such as Aurora~\cite{aurora}, GenCast~\cite{gencast}, and Earth-2~\cite{earth2} have been demonstrated to inform decisions during hurricanes and flooding events.  While most AI inference benchmarks tend to focus on throughput, in urgent computing scenarios the latency, or time to first result, is often equally (or more) important.  Some of the more recent MLPerf Inference\cite{mlperf,mlperfinference} benchmarks, such as the Edge\cite{mlperfedge} and Tiny\cite{mlperftiny} subsets, include single stream and multiple stream latency measurements in addition to throughput, which is a welcome addition for urgent use cases.  The datacenter and HPC subsets, which are more relevant for HPC scenarios, include only throughput, but could potentially be extended to include latency metrics.

\subsection{User Training}

Interactive and urgent HPC capabilities attract both traditional and unconventional users such as engineering, chemistry, social sciences, and bioinformatics. Unconventional users often are not used to scaling up their simulations to run in parallel on more cores and/or compute nodes, nor are they familiar with the usual software tools and environments of HPC. Similarly traditional HPC users may not be as familiar with the interactive and urgent HPC environments, requiring time to become confident with the new techniques and tools that this involves. Consequently, an important question is how to best train these users with pertinent tools so they can quickly achieve success and be motivated to leverage these technologies. For unconventional users becoming proficient and confident in using HPC systems will likely seem especially daunting.  

There are a large number of HPC training resources freely available, and it is crucial that expert and patient computational scientists and engineers who work as research facilitators are available to pointing users to such content. Research facilitators should support users by both helping them solve their issues and also building up the knowledge and confidence to tackle similar problems effectively in the future.
For common questions and issues, boilerplate emails and online content (e.g., knowledge bases) can reduce the load on research facilitators, while getting users acclimated to searching and accessing online content~\cite{mullen2018lessons}.  
Furthermore consortium content, such as HPC Carpentry\footnote{\url{https://www.hpc-carpentry.org/}}, provides shared resources for teaching basic HPC skills. 
Another example is the HPC Certification Forum\footnote{\url{https://www.hpc-certification.org/}} which is an effort to identify and organize competencies that clearly define the competences for practitioners, trainers and learners. However, crucially urgent computing is not yet covered in this forum.
Some HPC centers rely on online courses to teach HPC basics including helping users configure their environment to take advantage of HPC capabilities~\cite{mullen2017learning}. 
Furthermore, some centers require completion of these online courses in order to obtain full default resource allocations. Such online courses can also be used to gain HPC certifications, which can be important for users in their career progression.

\subsection{User Case Studies}

Discussions of theory, policy, and technologies is quite important, but they may not convey the experiences of actually implementing them; that's where user case studies come in. 
Many case studies have been shared during the Interactive and Urgent HPC workshops, from astronomy~\cite{htet2025modeling} to geosciences~\cite{odaka2020pangeo}, and from multi-physics modeling~\cite{viot2023from} to disease progress modeling~\cite{brown2021utilizing}. We cannot not highlight all of them in this chapter because there are so many of them; we encourage readers to access the workshop proceedings to read more about them. However, we will share a few representative use cases here. 

\subsubsection{Example: AI Inference Services}


As LLMs and GenAI models have gained much attention, researchers are eager to run, examine, and explore a wide variety of such models. HPC systems are a clear choice for such activities, but running them must integrate well into HPC in a secure manner. One notable approach to securely hosting Large Language Models (LLMs) on existing HPC infrastructure is SAIA~\cite{Doosthosseini2025, deckerphd}, which powers the Chat AI system by GWDG. It provides a seamless, Slurm-native solution, for creating high-performance private LLM-based services. SAIA's architecture addresses the challenge of running real-time, interactive services on HPC systems that are traditionally designed for batch processing. There is a decoupling here between the web-facing front-end and the secure HPC back-end, using a hardened SSH connection as the communication channel between them. To bridge the gap between service and batch paradigms, a custom scheduler script runs on top of Slurm, managing a dynamic pool of LLM instances. This script handles demand-based scaling and performs load balancing, allowing the LLM service to operate alongside regular HPC workloads. Data privacy is a central tenet of the design, where user prompts and conversation histories are never stored on the server, and instead remain exclusively on the user's local device to ensure confidentiality. In its production deployment, the architecture introduces a negligible latency overhead of around 23 ms~\cite{Doosthosseini2025}, with the primary performance bottleneck being the LLM inference speed itself. The throughput of the SSH connection is approximately 200 requests per second, which is sufficient for the LLM service's needs. 

Another approach to accommodate a highly interactive service like AI inference on an HPC system is to deploy cloud environments with Kubernetes, as outlined in section \ref{sec:scheduling}. As an alternative to SAIA, GWDG has also developed an AI inference platform based on that paradigm. The Scalable Kubernetes Inference Platform(SKIP)~\cite{DMK25, EKDDARCUWD25, deckerphd} is a cloud-native architecture built upon Kubernetes and its mature open-source ecosystem. Kubernetes environments are dynamically deployed using Warewulf~\cite{warewulf} by having compute nodes boot into stateless OS images that come pre-configured with Kubernetes dependencies and an automated installation script. 

For its own production environment. GWDG chose the Slurm and HPC native solution SAIA over Kubernetes based SKIP for serving AI inference. SKIP heavily relies on Kubernetes and its mature open-source ecosystem. The advantage is that updates and bug fixes are handled by active open-source communities. However, it requires extensive administrative expertise to maintain these complex technologies and suffers from severe vendor lock-in to the Kubernetes ecosystem. SAIA, on the other hand, has a much smaller dependency footprint, allowing more control over the platform. SKIP has shown performance overhead and latency bottlenecks due to the complex network virtualization layers of Kubernetes and its reliance on SSH tunneling between the cloud and HPC environments. SAIA has less architectural overhead because it avoids deep network virtualization~\cite{deckerphd}.

\subsubsection{Example: HPC In the Loop for AI and Digital Twins}


\begin{figure}
    \centering
    \includegraphics[width=0.95\textwidth]{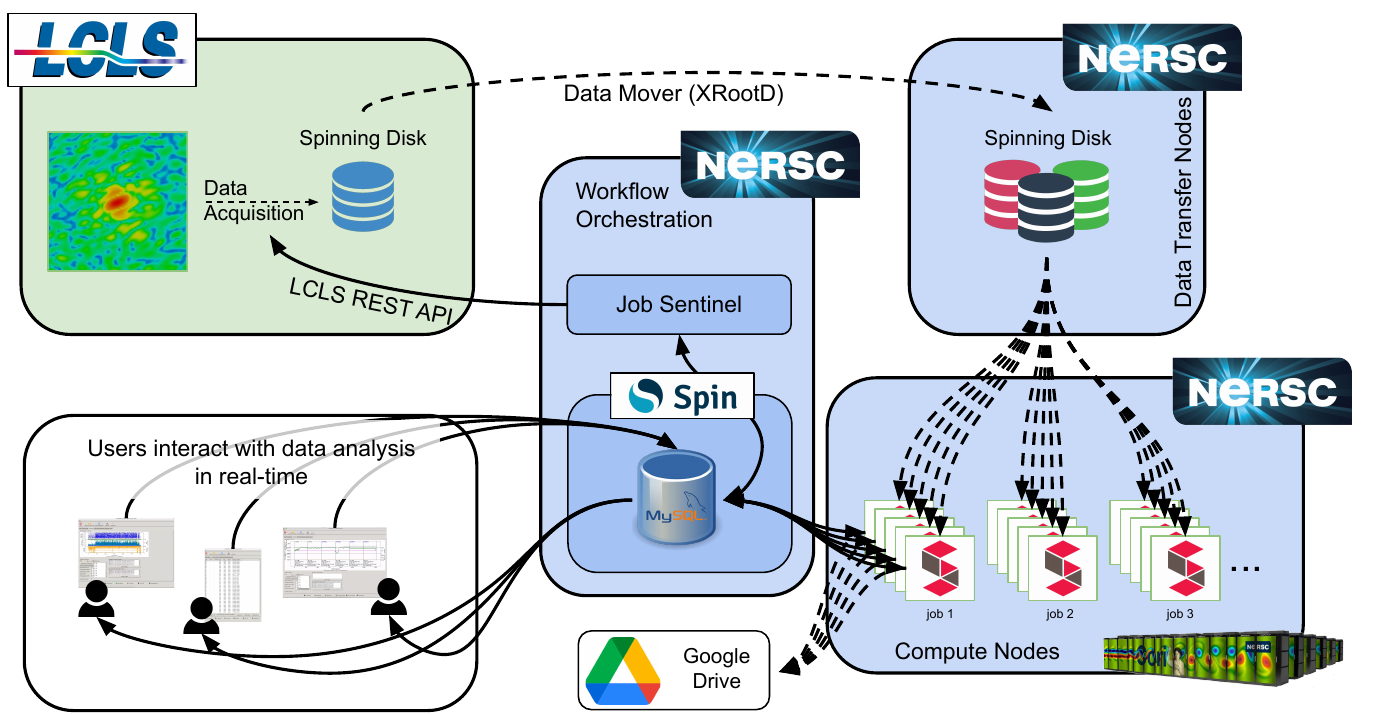}
    \caption{Workflow diagram of the {\it CCTBX.xfel} workflow manager. Colored
    boxes represent different sites at which the workflow is executed: i) The
    green box represents data collection at LCLS; ii) Blue boxes represent
    workflow components hosted at NERSC; iii) white boxes represent the rest of
    the world. Dashed arrows represent the flow of data: a) Raw data files are
    automatically transferred from ``fast feedback'' storage system at LCLS to
    the Lustre storage system at NERSC; b) Data moves from the NERSC Lustre
    storage system to on-node memory via the HPC system's high-speed network; c)
    analysis artifacts are automatically uploaded to Google Drive, to be shared
    by the research team. Solid lines represent flow of control. 1) Central to
    the whole workflow is the {\it CCTBX.xfel} Graphical User Interface (GUI)
    which gives users a realtime overview of data in motion, and completed and
    ongoing data analysis jobs (white box, bottom left); 2.) Workflow state is
    held in a {\it MySQL} database hosted on NERSC's k8s microservices platfrom
    (Spin -- cental box); 3.) New data -- or data in flight -- is queried using
    the LCLS REST API; 4.) A ``Job Sentinel'' continuously compares data sets,
    running and completed jobs, and data analysis parameters in order to
    determine which analyses need to be (re)run. {\it Image based on
    \cite{giannakou2021experiences}, and modified with the permission of the
    authors}.}
    \label{fid:cctbx_1}
\end{figure}

The Linac Coherent Light Source (LCLS) at SLAC is the world's most powerful
X-ray Free Electron Laser (XFEL). A pressing challenge with conducting science
at XFEL light sources is due to these instruments being an extremely powerful
yet scarce resource. This behooves instrument operators and researchers to
attempt to ``make every shot count'' in order to avoid wasting precious
instrument time. Here we explore Serial Femtosecond Crystallography (SFX) at the
LCLS XFELs as a case study~\cite{giannakou2021experiences, blaschke2023srn,
blaschke2021realtime}.

A common problem in SFX studies is how to infer scientifically-interesting
parameters (such as the state of chemical bonds) from a signal which also
contains random noise (such as crystal defects). To solve this problem, the LCLS
constructs a digital twin using the data that is collected in real time. For
SFX-type analyses this requires iterating on a model of the atomic configuration
of the sample comparing simulated sensor output with measurements at the same
time that those measurements are taking place. Fig.~\ref{fid:cctbx_1} gives a
general outline of the LCLS digital twin using the ExaFEL software
suite~\cite{blaschke2024exafel}.

The workflow described here is emblematic not only of SFX studies at XFELs but a
broader range of experimental and discovery science workflows requiring seamless
highly responsive large-s HPC \cite{etz2025enabling, blaschke2024exafel}.
A common requirement of similar workflows is that the HPC software stack is
portable. This way users can take advantage of as many HPC data centers as
possible, increasing the likelihood of finding available HPC resources which
coincide with scheduled measurement time at scientific instruments. In the case
of the {\it CCTBX.xfel} workflow shown in Figure~\ref{fid:cctbx_1} the entire
software stack is containerized; the workflow orchestrator was deployed on
k8s -- a common microservices platform which is available at many HPC data
centers and which can be deployed in userspace (under certain condition) for
simple workloads -- and performance portability was accomplished using the Kokkos
programming model\cite{wittwer2024ray-tracing, kokkos}.

This investment into workflow portability was put to the test, when an
unexpected power outage at the Lawrence Berkeley National Laboratory caused the
NERSC data center to lose power to the majority of its HPC systems in 2025.
This event coincided with an experiment at LCLS which was actively running {\it
CCTBX.exfel}. The workflow was interrupted for approximately one hour
(corresponding to about three samples), during which the experiment at LCLS was
``flying blind''. At NERSC the k8s cluster maintains continuous up-time during
power outages. This allowed the database to be migrated to the Stanford Shared
Data Facility (S3DF) and the Oak Ridge Leadership Computing Facility (OLCF) using
the {\it mysqldump} utility. Approximately 30 minutes after the power failure,
all raw data and database files were migrated to OLCF, and the {\it
CCTBX.exfel} workflow started to be deployed there also with realtime
``production'' jobs picking up where data analysis had left off about one hour
after the power failure.

\subsubsection{Example: Scheduling Challenges of Integrating Quantum Computing and HPC}

Integrating quantum computing resources with HPC is crucial for exploiting the potential of hybrid quantum-classical algorithms~\cite{beck/j.future.2024.06.058}, which are the most promising path to a practical quantum advantage on near-term quantum systems~\cite{roca2021}. This integration poses scheduling challenges similar to those of urgent computing, as, once a quantum computing device becomes available, it needs HPC resources with immediacy to allow the realization of an advantage~\cite{honda2025advantages}. This requirement is exacerbated in the case of mid-circuit measurements~\cite{Shehata_2026}, which need to complete classical HPC parts before the quantum circuit state de-coheres. Indeed the HPC community's growing interest in quantum and recent integration of HPC and quantum computing, for example in JUNIQ at the Jülich Supercomputing Centre, will potentially \emph{force the community's hand} and drive a move away from the traditional way of working. 

Several approaches have been discussed to co-schedule HPC resources and quantum devices under these boundary conditions. For example, external workflow managers such as StreamFlow or NextFlow can be used to ensure timely availability of necessary resources~\cite{viviani2025elephant}. Slurm heterogeneous jobs (hetjobs) can be used to reduce the idle time of the HPC partition, as it enables Slurm to schedule classical tasks according to the availability of the quantum device for the corresponding quantum tasks~\cite{esposito2023hybrid}. It has also been proposed to extend the Slurm hetjob capability with a credit system to throttle access to a quantum device, thereby allowing for more efficient scheduling of the required HPC resources~\cite{shehata2025bridging}.

Another approach to tackle these issues is resource malleability~\cite{viviani2025elephant, rocco2025dynamic} where the workflow of a hybrid application is governed by the HPC partition using an MPI application. Once a task is offloaded to a quantum device, the Dynamic Management of Resources (DMR) framework is used to minimize the used HPC resources to a single MPI process and release unused HPC resources back to the workload manager. After returning from the quantum device, the MPI application is expanded again corresponding to the computational demand of the upcoming classical task. Whilst this represents a significant departure from the current way of working, it has been demonstrated that this approach compares favorably with reserving a fixed sized HPC partition for the entire runtime of a hybrid application both in terms of time to completion (-44\%) and resource usage (-54\%) under resource contention. Indeed, this malleability also still compares favorably to using a workflow to allocate HPC resources on demand in terms of completion time (-8\%), albeit using more resources (+17.5\%)~\cite{rocco2025dynamic}.


 


\section{The Path Forward}
\label{sec:pathforward}

Ubiquitous laptops, tablets, smartphones, and cloud computing have ushered in a much greater expectation for interactive and urgent computing. 
Faculty, research staff, and students at universities around the world are gaining access to HPC systems through interactive HPC portals and research desktops hosted on HPC systems. We are seeing the next generation of scientists, engineers, mathematicians, economists, social scientists, etc. are learning how to use HPC systems by way of interactive and urgent HPC tools -- the next generation is being trained on interactive and urgent HPC capabilities! And present and emerging areas including data science, AI/ML, and quantum computing will rely more on interactive and urgent HPC in the coming years. The opportunities are exciting and abundant! 

But there is still plenty of research, prototyping, and development to be done. For instance: 
\begin{itemize}

\item Finding the balance between shorter, interactive jobs versus urgent, data streaming jobs versus larger, long-running simulations continues to be a challenge. Greater understanding, not only of the tradeoffs, but also of the metrics, evaluations, and best practices would help most HPC centers better serve their userbase. This applies to both organizational policies and scheduler policies. 

\item Open OnDemand, JupyterHub, proxy services, remote desktop servers for GUIs, and web dashboards today give users access to HPC, bypassing the Linux command line that many researchers find unfamiliar, intimidating, or just not the right fit for scientific workflows. 
Extrapolating that this trend of bringing familiar user interface paradigms to HPC will continue, we expect that future researchers will expect to interact with HPC through tablets or phones, and through new modalities like natural language.
This opens up new opportunities to make HPC more usable and accessible, but will require HPC developers to center on user experience and human-computer interaction --- a new area of research for HPC.

\item Today's researchers grew up with real-time collaborative productivity tools like Google Docs, but the closest analogy in a typical HPC context is account sharing --- entirely incompatible with HPC security models.
How can the needs and expectations of users be reconciled with HPC center infrastructure, networking, and security requirements?

\item While HPC infrastructure is capable of handling large volumes of raw data, more research, development, and improvements to center policy are needed in order to allow urgent and interactive HPC users to effectively and reliably deploy their workflows on HPC. These efforts include reliable high-speed data deliver; effective data transport through streaming; and collaborative data streaming and data storage access models. 

\item The supercomputing community has developed a rich suite of benchmarks with which to better understand the performance capabilities of their systems, but there is plenty of opportunity to extend these benchmarks and metrics to issues and challenges that interactive and urgent HPC users face. 

\item Typically HPC users are trained to use a given HPC system with either online documentation and/or classroom style tutorials in the classroom or over an online meeting. But some HPC centers have been developing and deploying asynchronous online courses (MOOCs) for such training. Which of these is most effective for students versus professionals who are learning to use the system? 

\item Looking even further ahead, can automated user support (like bots or HPC-Clippy!) help inexperienced users get up to speed quickly? How effectively can automated user support suggest the next steps for getting an interactive or urgent job running, and how complex of a situation can such automated user support handle? And do the users even like having such automated support to help them use the system? 

\end{itemize}

As these examples illustrate, there are many opportunities for research, prototyping, and development of interactive, urgent, real-time HPC for the HPC community to pursue~\cite{reuther2024interactiveurgenthpcchallenges}. 


\section{Conclusion}
\label{sec:conclusion}

In this chapter we have outlined the current state of the art in interactive and urgent capabilities on HPC. This has touched on policy, technology, scheduling, benchmarking, education, training, and user case studies, and indeed all these areas must be addressed in order to find the most appropriate paths forward. We have highlighted the potential importance of interactivity and urgent computing for the HPC community, and given societal and technological pressures, it is in our opinion highly likely that HPC will be forced to adopt this way of working regardless.

The case for increased consideration and development of interactive and urgent capabilities in HPC is compelling. Our position is rooted in first-hand experience, observation, and scholarship at the interface between HPC and high-impact, time-sensitive workflows in science and engineering. From the areas surveyed in this chapter several priorities and recommendations have emerged:

\begin{itemize}
    \item  
    \textbf{Urgency or time-sensitivity is an indicator of value.} As legendary computer scientist Grace Hopper\footnote{\url{https://www.youtube.com/watch?v=si9iqF5uTFk}} already mentioned in 1982, not all data are created equal and their value is highest shortly after creation. Imagine pointing your telescopes at the right point in the sky when the beginning of a supernova event was detected, or imagine precious beamtime at a user facility (e.g., a light source) where the lack of computed feedback is akin to running your experiment "blind". These scientific enterprises dedicate tremendous resources equal to or exceeding that of HPC centers to generate the data. The HPC community should treat their workflows with the same dedication, with priority on the scheduler (or even idle resources), priority on the network, and priority on storage.
    \item 
    \textbf{Unified infrastructures capable of handling both batch and interactive or urgent workloads are the way forward.} Separating scheduling and infrastructure for interactive and cloud-like workloads from those for classical batch processing leads to less flexibility in meeting changing hardware and software requirements, missed opportunities in balancing varying loads, overhead in operating separate software stacks, and reduced efficiency due to fragmentation into smaller hardware islands. Unified infrastructure approaches solve these problems for the manageable price of increased complexity of the combined stack. Whether unification is pursued by enabling cloud ecosystems to efficiently handle batch operation — for example by having Kubernetes manage HPC partitions — or by enabling classical HPC infrastructures to handle cloud-like workloads — for example by adding service management capabilities to Slurm — is open to debate and depends on a center's specific expertise and requirements. 
    \item \textbf{Streaming data into compute should become standard operating practice.} While HPC facilities often provide dedicated compute resources to their users, many other components (network, interconnect, storage) are still shared based on fair use. Noisy neighbors on shared resources can easily make the resource's performance characteristics to crumble causing huge tail latencies for everyone else. An example would be off-platform storage that is under stress from a workflow that touches millions of small files. Many teams are therefore switching to streaming workflows where the data is routed over the network directly into the compute nodes. Even though streaming is a component of every file transfer, there is currently no standardized tooling or guidelines that would make this switch easier and operate at scale. Users are left to find their own implementations. This needs to change. Ideally, the user community converges on a few tools or paradigms to better facilitate switching for those interested, and HPC facilities adjust their policies to make data streaming a first class citizen of getting scientific data into compute nodes.
\end{itemize}

Ultimately it is in the tackling, studying, learning, and sharing about advances in these challenges and opportunities that our community will go forth to have greater impact on scientific, societal and technical projects that require interactive and urgent HPC. 

\section*{Acknowledgments}

DISTRIBUTION STATEMENT A. Approved for public release. Distribution is unlimited.
This material is based upon work supported by the Department of the Air Force under Air Force Contract No. FA8702-15-D-0001
or FA8702-25-D-B002. Any opinions, findings, conclusions or recommendations expressed in this material are those of the
author(s) and do not necessarily reflect the views of the Department of the Air Force.

This work was supported by the Director, Office of Science, Office of Advanced Scientific Computing Research, of the U.S. Department of Energy under Contract No.~DE-AC02-05CH11231.

This work was supported by the Bundesministerium für Bildung und Forschung and the Niedersächsisches Ministerium für Wissenschaft und Kultur (KISSKI, Förderkennzeichen: 01IS22093A, and Nationales Hochleistungsrechnen an Hochschulen, NHR).

\bibliographystyle{unsrt}  
\bibliography{references}  

@article{swelborn2024mam,
    author = {Welborn, Samuel S and Harris, Chris and Ribet, Stephanie M and Varnavides, Georgios and Ophus, Colin and Enders, Bjoern and Ercius, Peter},
    title = {Streaming Large-Scale Microscopy Data to a Supercomputing Facility},
    journal = {Microscopy and Microanalysis},
    volume = {31},
    number = {1},
    pages = {ozae109},
    year = {2024},
    month = {11},
    issn = {1431-9276},
    doi = {10.1093/mam/ozae109},
    url = {https://doi.org/10.1093/mam/ozae109},
    eprint = {https://academic.oup.com/mam/article-pdf/31/1/ozae109/60676845/ozae109.pdf},
}

@article{blaschke2023srn,
	author = {Johannes P. Blaschke, Felix Wittwer, Bjoern Enders and Debbie Bard},
	doi = {10.1080/08940886.2023.2245700},
	eprint = {https://doi.org/10.1080/08940886.2023.2245700},
	journal = {Synchrotron Radiation News},
	number = {0},
	pages = {1-7},
	publisher = {Taylor & Francis},
	title = {How a Lightsource Uses a Supercomputer for Live Interactive Analysis of Large Data Sets},
	url = {https://doi.org/10.1080/08940886.2023.2245700},
	volume = {0},
	year = {2023}
}

@ARTICLE{rao2020deluge,
  title     = "Synchrotrons face a data deluge",
author = {Rao, Rahul},
  journal   = "Phys. Today",
  publisher = "AIP Publishing",
  volume    =  2020,
  number    =  2,
  pages     = "0925a",
  month     =  sep,
  year      =  2020,
  language  = "en"
}

@ARTICLE{Spurgeon2021-ym,
  title    = "Towards data-driven next-generation transmission electron
              microscopy",
  author   = "Spurgeon, Steven R and Ophus, Colin and Jones, Lewys and
              Petford-Long, Amanda and Kalinin, Sergei V and Olszta, Matthew J
              and Dunin-Borkowski, Rafal E and Salmon, Norman and Hattar,
              Khalid and Yang, Wei-Chang D and Sharma, Renu and Du, Yingge and
              Chiaramonti, Ann and Zheng, Haimei and Buck, Edgar C and Kovarik,
              Libor and Penn, R Lee and Li, Dongsheng and Zhang, Xin and
              Murayama, Mitsuhiro and Taheri, Mitra L",
  journal  = "Nature materials",
  volume   =  20,
  number   =  3,
  pages    = "274--279",
  month    =  mar,
  year     =  2021,
  language = "en",
  issn     = "1476-1122, 1476-4660",
  doi      = "10.1038/s41563-020-00833-z"
}

@article{wittwer2024ray-tracing,
author = {Wittwer, Felix and Sauter, Nicholas K. and Mendez, Derek and Poon, Billy K. and Brewster, Aaron S. and Holton, James M. and Wall, Michael E. and Hart, William E. and Bard, Deborah J. and Blaschke, Johannes P.},
title = {Accelerating x-ray tracing for exascale systems using Kokkos},
journal = {Concurrency and Computation: Practice and Experience},
volume = {36},
number = {5},
pages = {e7944},
doi = {https://doi.org/10.1002/cpe.7944},
url = {https://onlinelibrary.wiley.com/doi/abs/10.1002/cpe.7944},
eprint = {https://onlinelibrary.wiley.com/doi/pdf/10.1002/cpe.7944},
year = {2024}
}

@article{kokkos,
  title = "Kokkos: Enabling manycore performance portability through polymorphic memory access patterns ",
  journal = "Journal of Parallel and Distributed Computing ",
  volume = "74",
  number = "12",
  pages = "3202 - 3216",
  year = "2014",
  note = "Domain-Specific Languages and High-Level Frameworks for High-Performance Computing ",
  issn = "0743-7315",
  doi = "https://doi.org/10.1016/j.jpdc.2014.07.003",
  url = "http://www.sciencedirect.com/science/article/pii/S0743731514001257",
  author = "H. Carter Edwards and Christian R. Trott and Daniel Sunderland"
}

@article{blaschke2024exafel, 
  year     = {2024}, 
  title    = {ExaFEL: extreme-scale real-time data processing for X-ray free electron laser science}, 
  author   = {Blaschke, Johannes P. and Bolotovsky, Robert and Brewster, Aaron S. and Donatelli, Jeffrey and DuJardin, Antoine and Feng, Wu-chun and Ganapati, Vidya and Kroeger, Wilko and Mendez, Derek and McCorquodale, Peter and Mirchandaney, Seema and O'Grady, Christopher P. and Paley, Daniel W. and Perazzo, Amedeo and Poitevin, Frederic P. and Poon, Billy K. and Ramakrishnaiah, Vinay B. and Sauter, Nicholas K. and Shah, Niteya and Slaughter, Elliott and Sweeney, Christine and Tchoń, Daniel and Uervirojnangkoorn, Monarin and Wittwer, Felix and Wall, Michael E. and Yoon, Chun Hong and Young, Iris D.}, 
  journal  = {Frontiers in High Performance Computing}, 
  doi      = {10.3389/fhpcp.2024.1414569}, 
  pmid     = {41170299}, 
  pmcid    = {PMC12570208}, 
  pages    = {1414569}, 
  volume   = {2}
}

@article{flatken2023vestec,
  title={Vestec: Visual exploration and sampling toolkit for extreme computing},
  author={Flatken, Markus and Podobas, Artur and Fellegara, Riccardo and Basermann, Achim and Holke, Johannes and Knapp, David and Kontak, Max and Krullikowski, Christian and Nolde, Michael and Brown, Nick and others},
  journal={IEEE Access},
  volume={11},
  pages={87805--87834},
  year={2023},
  publisher={IEEE}
}

@inproceedings{gibb2020bespoke,
  title={A bespoke workflow management system for data-driven urgent hpc},
  author={Gibb, Gordon PS and Brown, Nick and Nash, Rupert W and Mendes, Miguel and Monedero, Santiago and Fidalgo, Humberto D{\'\i}az and Cisneros, Joaqu{\'\i}n Ram{\'\i}rez and Cardil, Adri{\'a}n and Kontak, Max},
  booktitle={2020 IEEE/ACM HPC for Urgent Decision Making (UrgentHPC)},
  pages={10--20},
  year={2020},
  organization={IEEE}
}

@article{brown2024predicting,
  title={Predicting accurate batch queue wait times on production supercomputers by combining machine learning techniques},
  author={Brown, Nick and Gibb, Gordon and Belikov, Evgenij and Nash, Rupert},
  journal={Concurrency and Computation: Practice and Experience},
  volume={36},
  number={15},
  pages={e8112},
  year={2024},
  publisher={Wiley Online Library}
}

@inproceedings{whitton2025we,
  title={Are we there yet? Predicting the queue wait times for hpc jobs},
  author={Whitton, Christin and Jones, William and Walker, Craig and Job, Vanessa and Senator, Steven and DeBardeleben, Nathan},
  booktitle={2025 IEEE International Conference on Cluster Computing (CLUSTER)},
  pages={1--12},
  year={2025},
  organization={IEEE}
}

@article{turilli2018comprehensive,
  title={A comprehensive perspective on pilot-job systems},
  author={Turilli, Matteo and Santcroos, Mark and Jha, Shantenu},
  journal={ACM Computing Surveys (CSUR)},
  volume={51},
  number={2},
  pages={1--32},
  year={2018},
  publisher={ACM New York, NY, USA}
}

@article{Zhou2020,
	author = {Zhou, Xiao-Yun and Guo, Yao and Shen, Mali and Yang, Guang-Zhong},
	date = {2020/08/01},
	doi = {10.1007/s11684-020-0770-0},
	id = {Zhou2020},
	isbn = {2095-0225},
	journal = {Frontiers of Medicine},
	number = {4},
	pages = {417--430},
	title = {Application of Artificial Intelligence in Surgery},
	url = {https://doi.org/10.1007/s11684-020-0770-0},
	volume = {14},
	year = {2020}}

@phdthesis{choy2002matlab,
  title={MATLAB* P 2.0: Interactive supercomputing made practical},
  author={Choy, Long Yin},
  year={2002},
  school={Massachusetts Institute of Technology}
}

@inproceedings{maeder2006gridmathematica,
  title={gridMathematica: overview and new developments},
  author={Maeder, Roman E},
  booktitle={Proceedings of the 2006 ACM/IEEE conference on Supercomputing},
  pages={252--es},
  year={2006}
}

@incollection{lin2018practitioner,
  title={Practitioner's Guide on the Use of Cloud Computing in Finance},
  author={Lin, Binghuan and Wehkamp, Rainer and Kanniainen, Juho},
  booktitle={High-Performance Computing in Finance},
  pages={509--536},
  year={2018},
  publisher={Chapman and Hall/CRC}
}

@ARTICLE{choy2005parallel,
  author={Choy, R. and Edelman, A.},
  journal={Proceedings of the IEEE}, 
  title={Parallel MATLAB: Doing it Right}, 
  year={2005},
  volume={93},
  number={2},
  pages={331-341},
  doi={10.1109/JPROC.2004.840490}
  }

@inproceedings{sakai2021experiences,
author = {Sakai, Scott and Mishin, Dmitry and Sivagnanam, Subhashini and Tatineni, Mahidhar and Kandes, Martin and Thomas, Mary and Irving, Christopher and Strande, Shawn and Norman, Michael},
title = {Experiences in Building a User Portal for Expanse Supercomputer},
year = {2021},
isbn = {9781450382922},
publisher = {Association for Computing Machinery},
address = {New York, NY, USA},
url = {https://doi.org/10.1145/3437359.3465590},
doi = {10.1145/3437359.3465590},
booktitle = {Practice and Experience in Advanced Research Computing 2021: Evolution Across All Dimensions},
articleno = {30},
numpages = {4},
location = {Boston, MA, USA},
series = {PEARC '21}
}

@ARTICLE{atwood2016secure,
  author={Atwood, Christopher A. and Goebbert, Randy C. and Calahan, Joshua A. and Hromadka III, Theodore V. and Proue, Thomas M. and Monceaux, Weston and Hirata, Jason},
  journal={Computing in Science \& Engineering}, 
  title={Secure Web-Based Access for Productive Supercomputing}, 
  year={2016},
  volume={18},
  number={1},
  pages={63-72},
  doi={10.1109/MCSE.2015.134}
}

@INPROCEEDINGS{prout2017mit,
  author={Prout, Andrew and Arcand, William and Bestor, David and Bergeron, Bill and Byun, Chansup and Gadepally, Vijay and Hubbell, Matthew and Houle, Michael and Jones, Michael and Michaleas, Peter and Milechin, Lauren and Mullen, Julie and Rosa, Antonio and Samsi, Siddharth and Reuther, Albert and Kepner, Jeremy},
  booktitle={2017 IEEE High Performance Extreme Computing Conference (HPEC)}, 
  title={MIT SuperCloud portal workspace: Enabling HPC web application deployment}, 
  year={2017},
  volume={},
  number={},
  pages={1-6},
  doi={10.1109/HPEC.2017.8091097}
}

@article{calegari2019web,
author = {Calegari, Patrice and Levrier, Marc and Balczy\'{n}ski, Pawe\l{}},
title = {Web Portals for High-performance Computing: A Survey},
year = {2019},
issue_date = {February 2019},
publisher = {Association for Computing Machinery},
address = {New York, NY, USA},
volume = {13},
number = {1},
issn = {1559-1131},
url = {https://doi.org/10.1145/3197385},
doi = {10.1145/3197385},
journal = {ACM Trans. Web},
month = feb,
articleno = {5},
numpages = {36}
}

@INPROCEEDINGS{byun2021node,
  author={Byun, Chansup and Arcand, William and Bestor, David and Bergeron, Bill and Gadepally, Vijay and Houle, Michael and Hubbell, Matthew and Jones, Michael and Klein, Anna and Michaleas, Peter and Milechin, Lauren and Mullen, Julie and Prout, Andrew and Reuther, Albert and Rosa, Antonio and Samsi, Siddharth and Yee, Charles and Kepner, Jeremy},
  booktitle={2021 IEEE High Performance Extreme Computing Conference (HPEC)}, 
  title={Node-Based Job Scheduling for Large Scale Simulations of Short Running Jobs}, 
  year={2021},
  volume={},
  number={},
  pages={1-7},
  doi={10.1109/HPEC49654.2021.9622870}}

@inproceedings{byun2020best,
   author = {Chansup Byun and Jeremy Kepner and William Arcand and David Bestor and Bill Bergeron and Vijay Gadepally and Michael Houle and Matthew Hubbell and Michael Jones and Andrew Kirby and Anna Klein and Peter Michaleas and Lauren Milechin and Julie Mullen and Andrew Prout and Antonio Rosa and Siddharth Samsi and Charles Yee and Albert Reuther},
   doi = {10.1109/HPEC43674.2020.9286142},
   isbn = {978-1-7281-9219-2},
   booktitle = {2020 IEEE High Performance Extreme Computing Conference (HPEC)},
   month = {9},
   pages = {1-7},
   publisher = {IEEE},
   title = {Best of Both Worlds: High Performance Interactive and Batch Launching},
   year = {2020},
}

@inbook{mullen2018lessons,
   author = {Julia Mullen and Albert Reuther and William Arcand and Bill Bergeron and David Bestor and Chansup Byun and Vijay Gadepally and Michael Houle and Matthew Hubbell and Michael Jones and Anna Klein and Peter Michaleas and Lauren Milechin and Andrew Prout and Antonio Rosa and Siddharth Samsi and Charles Yee and Jeremy Kepner},
   city = {Frankfurt},
   doi = {10.1007/978-3-030-02465-9_47},
   editor = {Rio Yokota and Michèle Weil and John Shalf and Sadaf Alam},
   journal = {ISC High Performance 2018: High Performance Computing},
   month = {6},
   pages = {655-668},
   publisher = {Springer, Cham},
   title = {Lessons Learned from a Decade of Providing Interactive, On-Demand High Performance Computing to Scientists and Engineers},
   url = {http://link.springer.com/10.1007/978-3-030-02465-9_47},
   year = {2018},
}

@article{mullen2017learning,
title = {Learning by doing, High Performance Computing education in the MOOC era},
journal = {Journal of Parallel and Distributed Computing},
volume = {105},
pages = {105-115},
year = {2017},
note = {Keeping up with Technology: Teaching Parallel, Distributed and High-Performance Computing},
issn = {0743-7315},
doi = {https://doi.org/10.1016/j.jpdc.2017.01.015},
url = {https://www.sciencedirect.com/science/article/pii/S0743731517300217},
author = {Julia Mullen and Chansup Byun and Vijay Gadepally and Siddharth Samsi and Albert Reuther and Jeremy Kepner}
}

@INPROCEEDINGS{reuther2005technology,
  author={Reuther, Albert I. and Currie, Tim and Kepner, Jeremy and Kim, Hahn G. and McCabe, Andrew and Michaleas, Peter and Travinin, Nadya},
  booktitle={2005 Users Group Conference (DOD-UGC'05)}, 
  title={Technology Requirements for Supporting On-Demand Interactive Grid Computing}, 
  year={2005},
  volume={},
  number={},
  pages={320-327},
  doi={10.1109/DODUGC.2005.65}
}

@article{reuther2006high,
   author = {Albert Reuther and Suzy Tichenor},
   issue = {4a},
   journal = {Cyberinfrastructure Technology Watch Quarterly},
   title = {High Performance Computing and Competitiveness: Making the Business Case},
   volume = {2},
   year = {2006},
}

@inproceedings{reuther2007technical,
   author = {Albert Reuther and Jeremy Kepner and Andy MCcabe and Julie Mullen and Nadya T. Bliss and Hahn Kim},
   booktitle = {Department of Defense - Proceedings of the HPCMP Users Group Conference 2007; High Performance Computing Modernization Program: A Bridge to Future Defense, DoD HPCMP UGC},
   pages = {403-409},
   title = {Technical challenges of supporting interactive HPC},
   year = {2007},
}

@inproceedings{reuther2004llgrid,
author="Albert Reuther and Tim Currie and Jeremy Kepner and Hahn Kim and Andrew McCabe and Michael P. Moore and Nadya Travinin",
title="LLGrid: Enabling on-demand grid computing with gridMatlab and pMatlab",
booktitle="High Performance Embedded Computing (HPEC) workshop",
year="2004",
pages="28--29"
}

@INPROCEEDINGS{jones2016scalability,
  author={Jones, Mike and Arcand, Bill and Bergeron, Bill and Bestor, David and Byun, Chansup and Milechin, Lauren and Gadepally, Vijay and Hubbell, Matt and Kepner, Jeremy and Michaleas, Pete and Mullen, Julie and Prout, Andy and Rosa, Tony and Samsi, Siddharth and Yee, Charles and Reuther, Albert},
  booktitle={2016 IEEE High Performance Extreme Computing Conference (HPEC)}, 
  title={Scalability of VM provisioning systems}, 
  year={2016},
  volume={},
  number={},
  pages={1-5},
  doi={10.1109/HPEC.2016.7761629}}

@INPROCEEDINGS{jones2018interactive,
  author={Jones, Michael and Kepner, Jeremy and Orchard, Bradley and Reuther, Albert and Arcand, William and Bestor, David and Bergeron, Bill and Byun, Chansup and Gadepally, Vijay and Houle, Michael and Hubbell, Matthew and Klein, Anna and Milechin, Lauren and Mullen, Julia and Prout, Andrew and Rosa, Antonio and Samsi, Siddharth and Yee, Charles and Michaleas, Peter},
  booktitle={2018 IEEE High Performance extreme Computing Conference (HPEC)}, 
  title={Interactive Launch of 16,000 Microsoft Windows Instances on a Supercomputer}, 
  year={2018},
  volume={},
  number={},
  pages={1-6},
  doi={10.1109/HPEC.2018.8547782}}

@INPROCEEDINGS{reuther2018interactive,
  author={Reuther, Albert and Kepner, Jeremy and Byun, Chansup and Samsi, Siddharth and Arcand, William and Bestor, David and Bergeron, Bill and Gadepally, Vijay and Houle, Michael and Hubbell, Matthew and Jones, Michael and Klein, Anna and Milechin, Lauren and Mullen, Julia and Prout, Andrew and Rosa, Antonio and Yee, Charles and Michaleas, Peter},
  booktitle={2018 IEEE High Performance extreme Computing Conference (HPEC)}, 
  title={Interactive Supercomputing on 40,000 Cores for Machine Learning and Data Analysis}, 
  year={2018},
  volume={},
  number={},
  pages={1-6},
  doi={10.1109/HPEC.2018.8547629}}

@article{reuther2018scalable,
   author = {Albert Reuther and Chansup Byun and William Arcand and David Bestor and Bill Bergeron and Matthew Hubbell and Michael Jones and Peter Michaleas and Andrew Prout and Antonio Rosa and Jeremy Kepner},
   doi = {10.1016/j.jpdc.2017.06.009},
   issn = {07437315},
   journal = {Journal of Parallel and Distributed Computing},
   title = {Scalable System Scheduling for HPC and Big Data},
   volume = {111},
   year = {2018},
}

@article{koehler2022,
title = {{Secure Authorization for RESTful HPC Access with FaaS Support}},
author = {K{\"o}hler, Christian and Biniaz, Mohammad Hossein and Bingert, Sven and Nolte, Hendrik and Kunkel, Julian},
year = {2022},
date = {2022},
journal = {{International Journal on Advances in Security}},
volume = {Vol. 15},
number = {No. 3 and 4},
pages = {119--131},
}

@INPROCEEDINGS{FirecREST,
	isbn = {978-0-7381-1055-4},
	doi = {10.1109/SuperCompCloud51944.2020.00009},
	year = {2020},
	booktitle = {2020 IEEE/ACM International Workshop on Interoperability of Supercomputing and Cloud Technologies (SuperCompCloud)},
	type = {Conference Paper},
	author = {Cruz, Felipe A. and Dabin, Alejandro J. and Dorsch, Juan P. and Koutsaniti, Eirini and Lezcano, Nelson F.},
	language = {en},
	address = {Piscataway, NJ},
	publisher = {IEEE},
	title = {FirecREST: A RESTful API to HPC systems},
	PAGES = {21 - 26},
}

@article{MaterialClouds,
	doi = {10.1038/s41597-020-00637-5},
	url = {https://doi.org/10.1038%2Fs41597-020-00637-5},
	year = 2020,
	month = {sep},
	publisher = {Springer Science and Business Media {LLC}},
	volume = {7},
	number = {1},
	author = {Leopold Talirz and Snehal Kumbhar and Elsa Passaro and Aliaksandr V. Yakutovich and Valeria Granata and Fernando Gargiulo and Marco Borelli and Martin Uhrin and Sebastiaan P. Huber and Spyros Zoupanos and Carl S. Adorf and Casper Welzel Andersen and Ole Schütt and Carlo A. Pignedoli and Daniele Passerone and Joost VandeVondele and Thomas C. Schulthess and Berend Smit and Giovanni Pizzi and Nicola Marzari},
	title = {Materials Cloud, a platform for open computational science},
	journal = {Scientific Data}
}

@article{AiiDA,
	doi = {10.1038/s41597-020-00638-4},
	url = {https://doi.org/10.1038%2Fs41597-020-00638-4},
	year = 2020,
	month = {sep},
	publisher = {Springer Science and Business Media {LLC}},
	volume = {7},
	number = {1},
	author = {Sebastiaan P. Huber and Spyros Zoupanos and Martin Uhrin and Leopold Talirz and Leonid Kahle and Rico Häuselmann and Dominik Gresch and Tiziano Müller and Aliaksandr V. Yakutovich and Casper W. Andersen and Francisco F. Ramirez and Carl S. Adorf and Fernando Gargiulo and Snehal Kumbhar and Elsa Passaro and Conrad Johnston and Andrius Merkys and Andrea Cepellotti and Nicolas Mounet and Nicola Marzari and Boris Kozinsky and Giovanni Pizzi},
	title = {{AiiDA} 1.0, a scalable computational infrastructure for automated reproducible workflows and data provenance},
	journal = {Scientific Data}
}

@article{OpenOnDemand, 
doi = {10.21105/joss.00622}, 
url = {https://doi.org/10.21105/joss.00622}, 
year = {2018}, 
publisher = {The Open Journal}, 
volume = {3}, 
number = {25}, 
pages = {622}, 
author = {Dave Hudak and Doug Johnson and Alan Chalker and Jeremy Nicklas and Eric Franz and Trey Dockendorf and Brian L. McMichael}, 
title = {Open OnDemand: A web-based client portal for HPC centers}, journal = {Journal of Open Source Software} 
}

@inproceedings{settlage2019open,
author = {Settlage, Robert and Chalker, Alan and Franz, Eric and Johnson, Doug and Gallo, Steve and Moore, Edgar and Hudak, David},
title = {Open OnDemand: HPC for Everyone},
year = {2019},
isbn = {978-3-030-34355-2},
publisher = {Springer-Verlag},
address = {Berlin, Heidelberg},
url = {https://doi.org/10.1007/978-3-030-34356-9_38},
doi = {10.1007/978-3-030-34356-9_38},
booktitle = {High Performance Computing: ISC High Performance 2019 International Workshops, Frankfurt, Germany, June 16-20, 2019, Revised Selected Papers},
pages = {504–513},
numpages = {10},
location = {Frankfurt, Germany}
}

@InProceedings{settlage2020tools,
author="Settlage, Robert
and Rajamohan, Srijith
and Lahmers, Kevin
and Chalker, Alan
and Franz, Eric
and Gallo, Steve
and Hudak, David",
editor="Juckeland, Guido
and Chandrasekaran, Sunita",
title="Portals for Interactive Steering of HPC Workflows",
booktitle="Tools and Techniques for High Performance Computing",
year="2020",
publisher="Springer International Publishing",
address="Cham",
pages="179--189",
isbn="978-3-030-44728-1"
}

@InProceedings{sadek2023open,
author="Sadek, Faras
and Munakami, Milson
and Barrett, Arthur
and Tan, Vesna
and Guillette, Jeremy
and Freeman, Robert M.
and Singh, Raminder",
editor="Bienz, Amanda
and Weiland, Mich{\`e}le
and Baboulin, Marc
and Kruse, Carola",
title="Open OnDemand Connector for Amazon Elastic Kubernetes Service (EKS)",
booktitle="High Performance Computing",
year="2023",
publisher="Springer Nature Switzerland",
address="Cham",
pages="577--586",
isbn="978-3-031-40843-4"
}

@inbook{HEAppE,
author = {Svaton, Vaclav and Martinovic, Jan and Křenek, Jan and Esch, Thomas and Tomancak, Pavel},
year = {2020},
month = {01},
pages = {280-293},
title = {HPC-as-a-Service via HEAppE Platform},
isbn = {978-3-030-22353-3},
doi = {10.1007/978-3-030-22354-0_26}
}

@InProceedings{Superfacility,
author="Bard, Deborah J.
and Day, Mark R.
and Enders, Bjoern
and Hartman--Baker, Rebecca J.
and Riney, John
and Snavely, Cory
and Torok, Gabor",
editor="Jagode, Heike
and Anzt, Hartwig
and Ltaief, Hatem
and Luszczek, Piotr",
title="Automation for Data-Driven Research with the NERSC Superfacility API",
booktitle="High Performance Computing",
year="2021",
publisher="Springer International Publishing",
address="Cham",
pages="333--345",
isbn="978-3-030-90539-2"
}

@INPROCEEDINGS{enders2020cross,
  author={Enders, Bjoern and Bard, Debbie and Snavely, Cory and Gerhardt, Lisa and Lee, Jason and Totzke, Becci and Antypas, Katie and Byna, Suren and Cheema, Ravi and Cholia, Shreyas and Day, Mark and Gaur, Aditi and Greiner, Annette and Groves, Taylor and Kiran, Mariam and Koziol, Quincey and Rowland, Kelly and Samuel, Chris and Selvarajan, Ashwin and Sim, Alex and Skinner, David and Thomas, Rollin and Torok, Gabor},
  booktitle={2020 IEEE/ACM 2nd Annual Workshop on Extreme-scale Experiment-in-the-Loop Computing (XLOOP)}, 
  title={Cross-facility science with the Superfacility Project at LBNL}, 
  year={2020},
  volume={},
  number={},
  pages={1-7},
  doi={10.1109/XLOOP51963.2020.00006}}

@misc{wilkinson2025designing,
  doi = {10.5281/ZENODO.17290392},
  url = {https://zenodo.org/doi/10.5281/zenodo.17290392},
  author = {Wilkinson, Sean and Widener, Patrick and Oral, Sarp and Ferreira da Silva, Rafael},
  language = {en},
  title = {Designing FAIR Workflows at OLCF: Building Scalable and Reusable Ecosystems for HPC Science},
  publisher = {Zenodo},
  year = {2025},  
  copyright = {Creative Commons Attribution 4.0 International}
}

@article{blaschke2021realtime, 
  year     = {2024}, 
  title    = {Real‐time XFEL data analysis at SLAC and NERSC: A trial run of nascent exascale experimental data analysis}, 
  author   = {Blaschke, Johannes P. and Brewster, Aaron S. and Paley, Daniel W. and Mendez, Derek and Bhowmick, Asmit and Sauter, Nicholas K. and Kröger, Wilko and Shankar, Murali and Enders, Bjoern and Bard, Deborah}, 
  journal  = {Concurrency and Computation: Practice and Experience}, 
  issn     = {1532-0626}, 
  doi      = {10.1002/cpe.8019}, 
  pmid     = {39935785}, 
  pmcid    = {PMC11809777}, 
  eprint   = {2106.11469}, 
  number   = {12}, 
  volume   = {36}
}

@article{vcluster,
  author       = {Maxime Martinasso and
                  Mark Klein and
                  Benjamin Cumming and
                  Miguel Gila and
                  Felipe A. Cruz and
                  Alberto Madonna and
                  Manuel Sopena Ballesteros and
                  Sadaf R. Alam and
                  Thomas C. Schulthess},
  title        = {Versatile Software-Defined Cluster for {HPC} Using Cloud Abstractions},
  journal      = {Comput. Sci. Eng.},
  volume       = {26},
  number       = {3},
  pages        = {20--29},
  year         = {2024},
  url          = {https://doi.org/10.1109/MCSE.2024.3394164},
  doi          = {10.1109/MCSE.2024.3394164},
  bibsource    = {dblp computer science bibliography, https://dblp.org}
}

@inproceedings{vncplay,
author = {Zeldovich, Nickolai and Chandra, Ramesh},
year = {2005},
pages = {189-198},
booktitle = {Proceedings of the FREENIX Track: 2005 USENIX Annual Technical Conference},
title = {Interactive Performance Measurement with {VNCPlay}.}
}

@INPROCEEDINGS{eithne,
  author={Jamieson, Maurice and Brown, Nick},
  booktitle={2020 IEEE/ACM HPC for Urgent Decision Making (UrgentHPC)}, 
  title={Benchmarking micro-core architectures for detecting disasters at the edge}, 
  year={2020},
  volume={},
  number={},
  pages={27-35},
  doi={10.1109/UrgentHPC51945.2020.00009}}

@unpublished{abaffy2009,
  title={Latencies in {Linux} and {FreeBSD} kernels with different schedulers – {O}(1), {CFS}, {4 BSD}},
  author={J. Abaffy and Tibor Krajcovic},
  year={2009},
  note={https://api.semanticscholar.org/CorpusID:10369071}
}

@INPROCEEDINGS{fan2020,
  author={Fan, Wei-Cong and Wong, Chee-Siang and Lee, Wai-Kong and Hwang, Seong-Oun},
  booktitle={2020 International Conference on Green and Human Information Technology (ICGHIT)}, 
  title={Comparison of Interactivity Performance of {L}inux {CFS} and {W}indows 10 {CPU} Schedulers}, 
  year={2020},
  volume={},
  number={},
  pages={31-34},
  doi={10.1109/ICGHIT49656.2020.00014}}

@misc{interbench,
    author = {C. Kolivas}, 
    title = {Interbench}, 
    note = {http://ck.kolivas.org/apps/interbench/}, 
    year = {2012}
}

@INPROCEEDINGS{wong2008,
  author={Wong, C.S. and Tan, I.K.T. and Kumari, R.D. and Lam, J.W. and Fun, W.},
  booktitle={2008 International Symposium on Information Technology}, 
  title={Fairness and interactive performance of {O}(1) and {CFS} {Linux} kernel schedulers}, 
  year={2008},
  volume={4},
  number={},
  pages={1-8},
  doi={10.1109/ITSIM.2008.4631872}}

@ARTICLE{mlperf,
  author={Mattson, Peter and Reddi, Vijay Janapa and Cheng, Christine and Coleman, Cody and Diamos, Greg and Kanter, David and Micikevicius, Paulius and Patterson, David and Schmuelling, Guenther and Tang, Hanlin and Wei, Gu-Yeon and Wu, Carole-Jean},
  journal={IEEE Micro}, 
  title={{MLPerf}: An Industry Standard Benchmark Suite for Machine Learning Performance}, 
  year={2020},
  volume={40},
  number={2},
  pages={8-16},
  doi={10.1109/MM.2020.2974843}}

@article{mlperfinference,
  title={{MLPerf} Inference Benchmark},
  author={Vijayarāghava Reḍḍī and Christina Miu Bing Cheng and David Kanter and Pete H Mattson and Guenther Schmuelling and Carole-Jean Wu and Brian Anderson and Maximilien Breughe and Mark Charlebois and William Chou and Ramesh Chukka and Cody A. Coleman and S. Davis and Pan Deng and Greg Diamos and Jared Duke and D. Fick and Julian Gardner and Itay Hubara and Sachin Satish Idgunji and Thomas B. Jablin and J. B. Jiao and Tom St. John and Pankaj Kanwar and David Lee and Jeffery Liao and Anton Lokhmotov and Francisco Massa and Peng Meng and Paulius Micikevicius and C. Kent Osborne and Gennady Pekhimenko and Arun Tejusve Raghunath Rajan and Dilip Sequeira and Ashish Sirasao and Fei Sun and Hanlin Tang and Michael Thomson and Frank Wei and Ephrem C. Wu and Ling Xu and Koichiro Yamada and Bing Yu and George Y. Yuan and Aaron Zhong and Pei Sheng Zhang and Yuchen Zhou},
  journal={2020 ACM/IEEE 47th Annual International Symposium on Computer Architecture (ISCA)},
  year={2019},
  pages={446-459},
  url={https://api.semanticscholar.org/CorpusID:207880425}
}

@misc{mlperfedge,
    author = {MLCommons}, 
    title = {{MLPerf} Inference: Edge}, 
    note = {https://mlcommons.org/benchmarks/inference-edge/}, 
    year = {2025}
}

@misc{mlperftiny,
    author = {MLCommons}, 
    title = {{MLPerf} Inference: Tiny}, 
    note = {https://mlcommons.org/benchmarks/inference-tiny/}, 
    year = {2025}
}

@misc{mlperfclient,
    author = {MLCommons}, 
    title = {{MLPerf} Client Benchmark}, 
    note = {https://mlcommons.org/benchmarks/client/}, 
    year = {2026}
}

@article{aurora,
author = {Bodnar, Cristian and Bruinsma, Wessel and Lucic, Ana and Stanley, Megan and Allen, Anna and Brandstetter, Johannes and Garvan , Patrick and Riechert, Maik and Weyn, Jonathan and Dong, Haiyu and Vaughan, Anna and Gupta, Jayesh and Thambiratnam, Kit and Archibald, Alex and Wu, Chun-Chieh and Heider, Elizabeth and Welling, Max and Turner, Richard and Perdikaris, Paris},
title = {A Foundation Model for the Earth System},
year = {2025},
month = {May},
url = {https://www.microsoft.com/en-us/research/publication/aurora-a-foundation-model-for-the-earth-system/},
pages = {1180-1187},
journal = {Nature},
volume = {641},
}

@INPROCEEDINGS{earth2,
       author = {{Kashinath}, Karthik and {Pritchard}, Michael and {Anandkumar}, Anima and {Pathak}, Jaideep and {Brenowitz}, Noah and {Cohen}, Yair and {Bonev}, Boris and {Messmer}, Peter and {Kurth}, Thorsten and {Azizzadenesheli}, Kamyar and {Subramaniam}, Akshay and {Geneva}, Nicholas and {Kovachki}, Nikola and {Kossaifi}, Jean and {Cherukuri}, Ram and {Hall}, David and {Choudhry}, Sanjay and {Posey}, Stan},
        title = "{NVIDIA Earth-2: Towards km-scale interactive digital twins}",
    booktitle = {AGU Fall Meeting Abstracts},
         year = 2023,
       volume = {2023},
        month = dec,
          eid = {IN52A-07},
        pages = {IN52A-07},
       adsurl = {https://ui.adsabs.harvard.edu/abs/2023AGUFMIN52A..07K}
}

@article{gencast,
    author = {Price, Ilan and Sanchez-Gonzalez, Alvaro and Alet, Ferran and Andersson, Tom R. and El-Kadi, Andrew and Masters, Dominic and Ewalds, Timo and Stott, Jacklynn and Mohamed, Shakir and Battaglia, Peter and Lam, Remi and Willson, Matthew},
    title = {Probabilistic weather forecasting with machine learning},
    year = {2025},
    month = {January},
    journal = {Nature},
    volume = {637},
    pages = {84-90},
}

@article{linpack,
  title={The {LINPACK} benchmark: past, present and future},
  author={Dongarra, Jack J and Luszczek, Piotr and Petitet, Antoine},
  journal={Concurrency and Computation: practice and experience},
  volume={15},
  number={9},
  pages={803--820},
  year={2003},
  url={https://doi.org/10.1002/cpe.728},
  doi={10.1002/cpe.728},
  publisher={Wiley Online Library}
}

@article{hpcg,
  title={High-performance conjugate-gradient benchmark: A new metric for ranking high-performance computing systems},
  author={Dongarra, Jack and Heroux, Michael A and Luszczek, Piotr},
  journal={The International Journal of High Performance Computing Applications},
  volume={30},
  number={1},
  pages={3--10},
  year={2016},
  url={https://doi.org/10.1177/1094342015593158},
  doi={10.1177/1094342015593158},
  publisher={SAGE Publications Sage UK: London, England}
}

@misc{graph500,
    author = {Graph 500}, 
    title = {Graph 500 | large-scale benchmarks}, 
    note = {https://graph500.org/}, 
    year = {2025}
}

@misc{green500,
    author={{TOP500.org}},
    title={Green500 | {TOP500}},
    note = {https://www.top500.org/lists/green500/},
    year = {2025}
}

@article{beck/j.future.2024.06.058,
author = {Beck, Thomas and Baroni, Alessandro and Bennink, Ryan and Buchs, Gilles and P\'{e}rez, Eduardo Antonio Coello and Eisenbach, Markus and da Silva, Rafael Ferreira and Meena, Muralikrishnan Gopalakrishnan and Gottiparthi, Kalyan and Groszkowski, Peter and Humble, Travis S. and Landfield, Ryan and Maheshwari, Ketan and Oral, Sarp and Sandoval, Michael A. and Shehata, Amir and Suh, In-Saeng and Zimmer, Christopher},
title = {Integrating quantum computing resources into scientific HPC ecosystems},
year = {2024},
issue_date = {Dec 2024},
publisher = {Elsevier Science Publishers B. V.},
address = {NLD},
volume = {161},
number = {C},
issn = {0167-739X},
url = {https://doi.org/10.1016/j.future.2024.06.058},
doi = {10.1016/j.future.2024.06.058},
journal = {Future Gener. Comput. Syst.},
month = dec,
pages = {11–25},
numpages = {15}
}

@article{roca2021,
  author       = {Cerezo de la Roca, Marco Vinicio Sebastian and Arrasmith, Andrew Thomas and Babbush, Ryan and Benjamin, Simon C. and Endo, Suguru and Fujii, Keisuke and McClean, Jarrod R. and Mitarai, Kosuke and Yuan, Xiao and Cincio, Lukasz and others},
  title        = {Variational quantum algorithms},
  doi          = {10.1038/s42254-021-00348-9},
  url          = {https://www.osti.gov/biblio/1880487},
  journal      = {Nature Reviews Physics},
  issn         = {ISSN 2522-5820},
  number       = {9},
  volume       = {3},
  place        = {United States},
  publisher    = {Springer Nature},
  year         = {2021},
  month        = {08}}

@misc{honda2025advantages,
      title={Advantages of Co-locating Quantum-HPC Platforms: A Survey for Near-Future Industrial Applications}, 
      author={Daigo Honda and Yuta Nishiyama and Junya Ishikawa and Kenichi Matsuzaki and Satoshi Miyata and Tadahiro Chujo and Yasuhisa Yamamoto and Masahiko Kiminami and Taro Kato and Jun Towada and Naoki Yoshioka and Naoto Aoki and Nobuyasu Ito},
      year={2025},
      eprint={2508.04171},
      archivePrefix={arXiv},
      primaryClass={quant-ph},
      url={https://arxiv.org/abs/2508.04171}, 
}

@article{Shehata_2026,
   title={Bridging paradigms: Designing for HPC-Quantum convergence},
   volume={174},
   ISSN={0167-739X},
   url={http://dx.doi.org/10.1016/j.future.2025.107980},
   DOI={10.1016/j.future.2025.107980},
   journal={Future Generation Computer Systems},
   publisher={Elsevier BV},
   author={Shehata, Amir and Groszkowski, Peter and Naughton, Thomas and Gopalakrishnan Meena, Muralikrishnan and Wong, Elaine and Claudino, Daniel and Ferreira da Silva, Rafael and Beck, Thomas},
   year={2026},
   month=jan, pages={107980} }

@misc{rocco2025dynamic,
      title={Dynamic Solutions for Hybrid Quantum-HPC Resource Allocation}, 
      author={Roberto Rocco and Simone Rizzo and Matteo Barbieri and Gabriella Bettonte and Elisabetta Boella and Fulvio Ganz and Sergio Iserte and Antonio J. Peña and Petter Sandås and Alberto Scionti and Olivier Terzo and Chiara Vercellino and Giacomo Vitali and Paolo Viviani and Jonathan Frassineti and Sara Marzella and Daniele Ottaviani and Iacopo Colonnelli and Daniele Gregori},
      year={2025},
      eprint={2508.04217},
      archivePrefix={arXiv},
      primaryClass={quant-ph},
      url={https://arxiv.org/abs/2508.04217}, 
}

@INPROCEEDINGS{viviani2025elephant,
  author={Viviani, Paolo and Rocco, Roberto and Barbieri, Matteo and Bettonte, Gabriella and Boella, Elisabetta and Cipollini, Marco and Frassineti, Jonathan and Ganz, Fulvio and Marzella, Sara and Ottaviani, Daniele and Rizzo, Simone and Scionti, Alberto and Vercellino, Chiara and Vitali, Giacomo and Terzo, Olivier and Montrucchio, Bartolomeo and Gregori, Daniele},
  booktitle={2025 55th Annual IEEE/IFIP International Conference on Dependable Systems and Networks Workshops (DSN-W)}, 
  title={Assessing the Elephant in the Room in Scheduling for Current Hybrid HPC-QC Clusters}, 
  year={2025},
  volume={},
  number={},
  pages={184-187},
  doi={10.1109/DSN-W65791.2025.00059}}

@INPROCEEDINGS{esposito2023hybrid,
  author={Esposito, Aniello and Jones, Jessica R. and Cabaniols, Sebastien and Brayford, David},
  booktitle={2023 IEEE International Conference on Quantum Computing and Engineering (QCE)}, 
  title={A Hybrid Classical-Quantum HPC Workload}, 
  year={2023},
  volume={02},
  number={},
  pages={117-121},
  doi={10.1109/QCE57702.2023.10194}}

@misc{shehata2025bridging,
      title={Bridging Paradigms: Designing for HPC-Quantum Convergence}, 
      author={Amir Shehata and Peter Groszkowski and Thomas Naughton and Murali Gopalakrishnan Meena and Elaine Wong and Daniel Claudino and Rafael Ferreira da Silvaa and Thomas Beck},
      year={2025},
      eprint={2503.01787},
      archivePrefix={arXiv},
      primaryClass={quant-ph},
      doi={https://doi.org/10.1016/j.future.2025.107980},
      url={https://arxiv.org/abs/2503.01787}, 
}

@misc{volcano,
    title = "Volcano",
    url = "https://volcano.sh",
    year = "2025"
}

@misc{mpi-operator,
    title = "MPI Operator",
    url = "https://github.com/kubeflow/mpi-operator",
    year = "2025"
}

@misc{soperator,
    title = "Soperator: Run Slurm in Kubernetes",
    url = "https://github.com/nebius/soperator",
    year = "2025"
}

@misc{sunk,
    title = "Introducing SUNK: A Slurm on Kubernetes Implementation for HPC and Large Scale AI",
    url = "https://coreweave.com/blog/sunk-slurm-on-kubernetes-implementations",
    year = "2025"
}

@misc{slurm-operator,
    title = "slurm-operator: Kubernetes Operator for Slurm Clusters",
    url = "https://github.com/SlinkyProject/slurm-operator",
    year = "2025"
}

@misc{slinky,
    title = "Slinky: Bridging Slurm And Kubernetes",
    url = "https://www.schedmd.com/slinky/why-slinky/",
    year = "2025"
}

@inproceedings{hpk,
author = {Chazapis, Antony and Vassilakis, Lefteris and Petsis, Giannis and Marazakis, Manolis and Bilas, Angelos},
title = {Evaluating HPK for Running Cloud-Native Workloads on Slurm Clusters},
year = {2025},
isbn = {9798400718717},
publisher = {Association for Computing Machinery},
address = {New York, NY, USA},
url = {https://doi.org/10.1145/3731599.3767352},
doi = {10.1145/3731599.3767352},
pages = {163–171},
numpages = {9},
location = {
},
series = {SC Workshops '25}
}

@article{Doosthosseini2025,
  title = {SAIA: A Seamless Slurm-Native Solution for HPC-Based Services},
  url = {http://dx.doi.org/10.21203/rs.3.rs-6648693/v1},
  DOI = {10.21203/rs.3.rs-6648693/v1},
  publisher = {Springer Science and Business Media LLC},
  author = {Doosthosseini,  Ali and Decker,  Jonathan and Nolte,  Hendrik and Kunkel,  Julian},
  year = {2025},
  month = jul 
}

@phdthesis{deckerphd,
  title = {Achieving Scalable AI Inference in a Unified Cloud and HPC Environment by Employing System of Systems Architecture Design},
  url = {http://dx.doi.org/10.53846/goediss-11674},
  DOI = {10.53846/goediss-11674},
  school = {University Goettingen},
  author = {Decker,  Jonathan}
}

@booklet{wells2019scheduling,
author = {Jack Wells},
title = {Scheduling Considerations for Leadership-Class Supercomputers},
howpublished= {Keynote talk at Second Workshop on Interactive High-Performance Computing at ISC-2019, Frankfurt, Germany},
year = {2019}, 
url = {https://www.interactivehpc.com/previous-editions/isc-2019}
}

@booklet{gradfreilich2023interactive,
author = {Silvina Grad-Frehlich},
title = {Interactive and Urgent Computing with MATLAB},
howpublished= {Keynote talk at Second Workshop on Interactive High-Performance Computing at ISC-2023, Hamburg, Germany},
year = {2023}, 
url = {https://www.interactivehpc.com/previous-editions/isc-2023}
}

@booklet{kunkel2024challenging,
author = {Julian Kunkel},
title = {A Challenging Transition to Offer Interactive and Urgent Computing at GWDG},
howpublished= {Keynote talk at the Third Combined Workshop on Interactive and Urgent High-Performance Computing at ISC-2024, Hamburg, Germany},
year = {2024}, 
url = {https://www.interactivehpc.com/previous-editions/isc-2024}
}

@article{raess2023teaching,
title = {Teaching Supercomputing and Software Engineering Skills to Science and Engineering Students},
journal = {ETH Learning and Teaching JournalEducational Development and Technology},
volume = {4},
number = {1},
pages = {22-35},
year = {2023},
doi = {https://doi.org/10.16906/lt-eth.v4i1.235},
author = {Ludovic Räss and Mauro A. Werder and Ivan Utkin and Samuel Omlin}
}

@inproceedings{zhukov2025onboarding,
author = {Zhukov, Ilya and Zjupa, Jolanta},
title = {Onboarding of HPC users: hands-on approach at Juelich Supercomputing Centre},
year = {2025},
isbn = {9798400713989},
publisher = {Association for Computing Machinery},
address = {New York, NY, USA},
url = {https://doi.org/10.1145/3708035.3736077},
doi = {10.1145/3708035.3736077},
booktitle = {Practice and Experience in Advanced Research Computing 2025: The Power of Collaboration},
articleno = {86},
numpages = {5},
location = {
},
series = {PEARC '25}
}

@InProceedings{sala2025novel,
author="Sala, Leonardo
and Talamo, Ivano
and Sharapov, Borys
and Assmann, Greta
and Dorigo, Alvise",
editor="Neuwirth, Sarah
and Paul, Arnab Kumar
and Weinzierl, Tobias
and Carson, Erin Claire",
title="A Novel Approach to Dynamic Computing Using Slurm",
booktitle="High Performance Computing",
year="2026",
publisher="Springer Nature Switzerland",
address="Cham",
pages="376--383",
isbn="978-3-032-07612-0"
}

@inproceedings{mcglothlin2025implementing,
author = {McGlothlin, Jay and Carothers, Christopher},
title = {Implementing support for Interactive and AI workloads in a traditional HPC environment},
year = {2025},
isbn = {9798400718717},
publisher = {Association for Computing Machinery},
address = {New York, NY, USA},
url = {https://doi.org/10.1145/3731599.3767473},
doi = {10.1145/3731599.3767473},
booktitle = {Proceedings of the SC '25 Workshops of the International Conference for High Performance Computing, Networking, Storage and Analysis},
pages = {2146–2150},
numpages = {5},
location = {
},
series = {SC Workshops '25}
}

@inproceedings{minami2025physical,
author = {Minami, Shohei and Endo, Toshio and Nomura, Akihiro and Ohtsuji, Hiroki and Kato, Jun and Miwa, Masahiro and Yoshida, Eiji},
title = {Physical System Study on Balancing Interactive and Batch Job Performance through Oversubscribing Scheduling},
year = {2025},
isbn = {9798400718717},
publisher = {Association for Computing Machinery},
address = {New York, NY, USA},
url = {https://doi.org/10.1145/3731599.3767472},
doi = {10.1145/3731599.3767472},
booktitle = {Proceedings of the SC '25 Workshops of the International Conference for High Performance Computing, Networking, Storage and Analysis},
pages = {2137–2145},
numpages = {9},
location = {
},
series = {SC Workshops '25}
}

@article{ahn2020flux,
title = {Flux: Overcoming scheduling challenges for exascale workflows},
journal = {Future Generation Computer Systems},
volume = {110},
pages = {202-213},
year = {2020},
issn = {0167-739X},
doi = {https://doi.org/10.1016/j.future.2020.04.006},
url = {https://www.sciencedirect.com/science/article/pii/S0167739X19317169},
author = {Dong H. Ahn and Ned Bass and Albert Chu and Jim Garlick and Mark Grondona and Stephen Herbein and Helgi I. Ingólfsson and Joseph Koning and Tapasya Patki and Thomas R.W. Scogland and Becky Springmeyer and Michela Taufer}
}

@InProceedings{henschel2024use,
author="Henschel, Robert
and Lindemann, Jonas
and Follin, Anders
and Dammann, Bernd
and Dennis, Cicada
and Thota, Abhinav",
editor="Weiland, Mich{\`e}le
and Neuwirth, Sarah
and Kruse, Carola
and Weinzierl, Tobias",
title="Use Cases for High Performance Research Desktops",
booktitle="High Performance Computing. ISC High Performance 2024 International Workshops",
year="2025",
publisher="Springer Nature Switzerland",
address="Cham",
pages="257--268",
isbn="978-3-031-73716-9"
}

@INPROCEEDINGS{lindeman2024interactive,
  author={Lindemann, Jonas and Follin, Anders},
  booktitle={SC24-W: Workshops of the International Conference for High Performance Computing, Networking, Storage and Analysis}, 
  title={Interactive HPC and the LUNARC Desktop Environment}, 
  year={2024},
  volume={},
  number={},
  pages={2012-2019},
  doi={10.1109/SCW63240.2024.00251}}

@InProceedings{welborn2024high-abe,
author="Welborn, Samuel S.
and Harris, Chris
and Ercius, Peter
and Bard, Deborah J.
and Enders, Bjoern",
editor="Weiland, Mich{\`e}le
and Neuwirth, Sarah
and Kruse, Carola
and Weinzierl, Tobias",
title="Accelerating Time-to-Science by Streaming Detector Data Directly into Perlmutter Compute Nodes",
booktitle="High Performance Computing. ISC High Performance 2024 International Workshops",
year="2025",
doi={10.1007/978-3-031-73716-9_17}, 
publisher="Springer Nature Switzerland",
address="Cham",
pages="243--256",
isbn="978-3-031-73716-9"
}

@inproceedings{maliaroudakis2022interactive,
author="Maliaroudakis, Evangelos
and Chazapis, Antony
and Kanterakis, Alexandros
and Marazakis, Manolis
and Bilas, Angelos",
editor="Anzt, Hartwig
and Bienz, Amanda
and Luszczek, Piotr
and Baboulin, Marc",
title="Interactive, Cloud-Native Workflows on HPC Using KNoC",
booktitle="High Performance Computing. ISC High Performance 2022 International Workshops",
year="2022",
publisher="Springer International Publishing",
address="Cham",
pages="221--232",
isbn="978-3-031-23220-6"
}

@article{luszczek2005introduction,
  title={Introduction to the HPC challenge benchmark suite},
  author={Luszczek, Piotr and Dongarra, Jack J and Koester, David and Rabenseifner, Rolf and Lucas, Bob and Kepner, Jeremy and McCalpin, John and Bailey, David and Takahashi, Daisuke},
  year={2005}
}

@INPROCEEDINGS{jamieson2020benchmarking,
  author={Jamieson, Maurice and Brown, Nick},
  booktitle={2020 IEEE/ACM HPC for Urgent Decision Making (UrgentHPC)}, 
  title={Benchmarking micro-core architectures for detecting disasters at the edge}, 
  year={2020},
  volume={},
  number={},
  pages={27-35},
  doi={10.1109/UrgentHPC51945.2020.00009}}

@INPROCEEDINGS{balouek2021evaluating,
  author={Balouek-Thomert, Daniel and Rodero, Ivan and Parashar, Manish},
  booktitle={2021 IEEE/ACM HPC for Urgent Decision Making (UrgentHPC)}, 
  title={Evaluating policy-driven adaptation on the Edge-to-Cloud Continuum}, 
  year={2021},
  volume={},
  number={},
  pages={11-20},
  doi={10.1109/UrgentHPC54802.2021.00007}}

@INPROCEEDINGS{hardy2021lessons,
  author={Hardy, David J. and Stone, John E. and Isralewitz, Barry and Tajkhorshid, Emad},
  booktitle={2021 IEEE/ACM HPC for Urgent Decision Making (UrgentHPC)}, 
  title={Lessons Learned from Responsive Molecular Dynamics Studies of the COVID-19 Virus}, 
  year={2021},
  volume={},
  number={},
  pages={1-10},
  doi={10.1109/UrgentHPC54802.2021.00006}}

@INPROCEEDINGS{cheng2021real,
  author={Cheng, Albert M. K.},
  booktitle={2021 IEEE/ACM HPC for Urgent Decision Making (UrgentHPC)}, 
  title={Real-Time COVID-19 Infection Risk Assessment and Mitigation based on Public-Domain Data}, 
  year={2021},
  volume={},
  number={},
  pages={29-35},
  doi={10.1109/UrgentHPC54802.2021.00009}}

@INPROCEEDINGS{mandel2019interactive,
  author={Mandel, Jan and Vejmelka, Martin and Kochanski, Adam and Farguell, Angel and Haley, James and Mallia, Derek and Hilburn, Kyle},
  booktitle={2019 IEEE/ACM HPC for Urgent Decision Making (UrgentHPC)}, 
  title={An Interactive Data-Driven HPC System for Forecasting Weather, Wildland Fire, and Smoke}, 
  year={2019},
  volume={},
  number={},
  pages={35-44},
  doi={10.1109/UrgentHPC49580.2019.00010}}

@INPROCEEDINGS{lovholt2019urgent,
  author={Løvholt, Finn and Lorito, Stefano and Macias, Jorge and Volpe, Manula and Selva, Jacopo and Gibbons, Steven},
  booktitle={2019 IEEE/ACM HPC for Urgent Decision Making (UrgentHPC)}, 
  title={Urgent Tsunami Computing}, 
  year={2019},
  volume={},
  number={},
  pages={45-50},
  doi={10.1109/UrgentHPC49580.2019.00011}}

@INPROCEEDINGS{goubier2020fast,
  author={Goubier, Thierry and Rakowsky, Natalja and Harig, Sven},
  booktitle={2020 IEEE/ACM HPC for Urgent Decision Making (UrgentHPC)}, 
  title={Fast Tsunami Simulations for a Real-Time Emergency Response Flow}, 
  year={2020},
  volume={},
  number={},
  pages={21-26},
  doi={10.1109/UrgentHPC51945.2020.00008}}

@InProceedings{farrell2018deep,
author="Farrell, Steve
and Vose, Aaron
and Evans, Oliver
and Henderson, Matthew
and Cholia, Shreyas
and P{\'e}rez, Fernando
and Bhimji, Wahid
and Canon, Shane
and Thomas, Rollin
and Prabhat",
editor="Yokota, Rio
and Weiland, Mich{\`e}le
and Shalf, John
and Alam, Sadaf",
title="Interactive Distributed Deep Learning with Jupyter Notebooks",
booktitle="High Performance Computing",
year="2018",
publisher="Springer International Publishing",
address="Cham",
pages="678--687",
isbn="978-3-030-02465-9"
}

@InProceedings{goebbert2018enabling,
author="G{\"o}bbert, Jens Henrik
and Kreuzer, Tim
and Grosch, Alice
and Lintermann, Andreas
and Riedel, Morris",
editor="Yokota, Rio
and Weiland, Mich{\`e}le
and Shalf, John
and Alam, Sadaf",
title="Enabling Interactive Supercomputing at JSC Lessons Learned",
booktitle="High Performance Computing",
year="2018",
publisher="Springer International Publishing",
address="Cham",
pages="669--677",
isbn="978-3-030-02465-9"
}

@InProceedings{henderson2020accelerating,
author="Henderson, Matthew L.
and Krinsman, William
and Cholia, Shreyas
and Thomas, Rollin
and Slaton, Trevor",
editor="Juckeland, Guido
and Chandrasekaran, Sunita",
title="Accelerating Experimental Science Using Jupyter and NERSC HPC",
booktitle="Tools and Techniques for High Performance Computing",
year="2020",
publisher="Springer International Publishing",
address="Cham",
pages="145--163",
isbn="978-3-030-44728-1"
}

@INPROCEEDINGS{ragan2024enabling,
  author={Ragan-Kelley, Min and Henderson, Matt and Pérez, Fernando and Thomas, Rollin and Cholia, Shreyas and Ramakrishnan, Lavanya},
  booktitle={SC24-W: Workshops of the International Conference for High Performance Computing, Networking, Storage and Analysis}, 
  title={Enabling Scientific Collaboration with JupyterHub}, 
  year={2024},
  volume={},
  number={},
  pages={1993-2002},
  doi={10.1109/SCW63240.2024.00249}}

@INPROCEEDINGS{werner2024jumper,
  author={Werner, Elias and Rygin, Anton and Gocht-Zech, Andreas and Döbel, Sebastian and Lieber, Matthias},
  booktitle={SC24-W: Workshops of the International Conference for High Performance Computing, Networking, Storage and Analysis}, 
  title={JUmPER: Performance Data Monitoring, Instrumentation and Visualization for Jupyter Notebooks}, 
  year={2024},
  volume={},
  number={},
  pages={2003-2011},
  doi={10.1109/SCW63240.2024.00250}}

@INPROCEEDINGS{giannakou2021experiences,
  author={Giannakou, Anna and Blaschke, Johannes P. and Bard, Deborah and Ramakrishnan, Lavanya},
  booktitle={2021 IEEE/ACM HPC for Urgent Decision Making (UrgentHPC)}, 
  title={Experiences with Cross-Facility Real-Time Light Source Data Analysis Workflows}, 
  year={2021},
  volume={},
  number={},
  pages={45-53},
  doi={10.1109/UrgentHPC54802.2021.00011}}

@InProceedings{etz2025enabling,
author="Etz, Brian D.
and Rogers, David M.
and Brim, Michael J.
and Maheshwari, Ketan
and Leland, Kellen
and Skluzacek, Tyler J.
and Lange, Jack
and Pelfrey, Daniel
and Webb, Jordan
and Widener, Patrick
and Adamson, Ryan
and Zimmer, Christopher
and Melesse Vergara, Ver{\'o}nica G.
and Shankar, Mallikarjun
and Oral, Sarp
and da Silva, Rafael Ferreira",
editor="Neuwirth, Sarah
and Paul, Arnab Kumar
and Weinzierl, Tobias
and Carson, Erin Claire",
title="Enabling Seamless Transitions from Experimental to Production HPC for Interactive Workflows",
booktitle="High Performance Computing",
year="2026",
publisher="Springer Nature Switzerland",
address="Cham",
pages="363--375",
isbn="978-3-032-07612-0"
}

@inproceedings{maheshwari2025evaluating,
author = {Maheshwari, Ketan and Borch, Anderson and Webb, Jordan and Etz, Brian and Miller, Ross and Suter, Fr\'{e}d\'{e}ric and Oral, Sarp and Ferreira da Silva, Rafael},
title = {Evaluating HPC Scheduling Strategies for Urgent Workloads},
year = {2025},
isbn = {9798400718717},
publisher = {Association for Computing Machinery},
address = {New York, NY, USA},
url = {https://doi.org/10.1145/3731599.3767474},
doi = {10.1145/3731599.3767474},
booktitle = {Proceedings of the SC '25 Workshops of the International Conference for High Performance Computing, Networking, Storage and Analysis},
pages = {2151–2160},
numpages = {10},
location = {
},
series = {SC Workshops '25}
}

@InProceedings{klein2020interactive,
author="Klein, Mark
and Martinasso, Maxime
and Leong, Siew Hoon
and Alam, Sadaf R.",
editor="Juckeland, Guido
and Chandrasekaran, Sunita",
title="Interactive Supercomputing for Experimental Data-Driven Workflows",
booktitle="Tools and Techniques for High Performance Computing",
year="2020",
publisher="Springer International Publishing",
address="Cham",
pages="164--178",
isbn="978-3-030-44728-1"
}

@inproceedings{htet2025modeling,
author = {Htet, Ye and Sudvarg, Marion and Yang, Honghao and Buhler, Jeremy and Chamberlain, Roger and Buckley, James},
title = {Modeling and Optimizing Real-Time Telescope Interaction for Multi-wavelength Observation of Gamma-ray Bursts},
year = {2025},
isbn = {9798400718717},
publisher = {Association for Computing Machinery},
address = {New York, NY, USA},
url = {https://doi.org/10.1145/3731599.3767475},
doi = {10.1145/3731599.3767475},
booktitle = {Proceedings of the SC '25 Workshops of the International Conference for High Performance Computing, Networking, Storage and Analysis},
pages = {2161–2168},
numpages = {8},
location = {
},
series = {SC Workshops '25}
}

@InProceedings{odaka2020pangeo,
author="Odaka, Tina Erica
and Banihirwe, Anderson
and Eynard-Bontemps, Guillaume
and Ponte, Aurelien
and Maze, Guillaume
and Paul, Kevin
and Baker, Jared
and Abernathey, Ryan",
editor="Juckeland, Guido
and Chandrasekaran, Sunita",
title="The Pangeo Ecosystem: Interactive Computing Tools for the Geosciences: Benchmarking on HPC",
booktitle="Tools and Techniques for High Performance Computing",
year="2020",
publisher="Springer International Publishing",
address="Cham",
pages="190--204",
isbn="978-3-030-44728-1"
}

@INPROCEEDINGS{brown2021utilizing,
  author={Brown, Nick and Nash, Rupert and Poletti, Piero and Guzzetta, Giorgio and Manica, Mattia and Zardini, Agnese and Flatken, Markus and Vidal, Jules and Gueunet, Charles and Belikov, Evgenij and Tierny, Julien and Podobas, Artur and Der Chien, Wei and Markidis, Stefano and Gerndt, Andreas},
  booktitle={2021 IEEE/ACM HPC for Urgent Decision Making (UrgentHPC)}, 
  title={Utilising urgent computing to tackle the spread of mosquito-borne diseases}, 
  year={2021},
  volume={},
  number={},
  pages={36-44},
  doi={10.1109/UrgentHPC54802.2021.00010}}

@InProceedings{viot2023from,
author="Viot, Louis
and Piel, Yannick
and Neumann, Philipp",
editor="Bienz, Amanda
and Weiland, Mich{\`e}le
and Baboulin, Marc
and Kruse, Carola",
title="From Desktop to Supercomputer: Computational Fluid Dynamics Augmented by Molecular Dynamics Using MaMiCo and preCICE",
booktitle="High Performance Computing",
year="2023",
publisher="Springer Nature Switzerland",
address="Cham",
pages="567--576",
isbn="978-3-031-40843-4"
}

@inproceedings{brown2019role,
author = {Brown, Nick and Nash, Rupert and Gibb, Gordon and Prodan, Bianca and Kontak, Max and Olshevsky, Vyacheslav and Der Chien, Wei},
title = {The Role of Interactive Super-Computing in Using HPC for Urgent Decision Making},
year = {2019},
isbn = {978-3-030-34355-2},
publisher = {Springer-Verlag},
address = {Berlin, Heidelberg},
url = {https://doi.org/10.1007/978-3-030-34356-9_40},
doi = {10.1007/978-3-030-34356-9_40},
booktitle = {High Performance Computing: ISC High Performance 2019 International Workshops, Frankfurt, Germany, June 16-20, 2019, Revised Selected Papers},
pages = {528–540},
numpages = {13},
location = {Frankfurt, Germany}
}

@misc{warewulf,
  title = {Warewulf/Warewulf},
  howpublished = {\url{https://github.com/warewulf/warewulf}},
  note = {Visited on July 7, 2025}
}

@article{EKDDARCUWD25,
	author	 = {Jonathan Decker and Julian Kunkel},
	title	 = {{Ephemeral Kubernetes: dynamically deleting and recreating clusters using Warewulf}},
	year	 = {2025},
	month	 = {10},
	journal	 = {The Journal of Supercomputing},
	number	 = {81},
	doi	 = {https://doi.org/10.1007/s11227-025-07668-y},
	url	 = {http://dx.doi.org/10.1007/s11227-025-07668-y},
}

@inproceedings{DMK25,
  author = {Decker, Jonathan and Metje, S\"{o}ren and Kunkel, Julian},
  title = {Running Kubernetes Workloads on Rootless HPC Systems Using Slurm},
  booktitle = {CLOUD COMPUTING 2025, The Sixteenth International Conference on Cloud Computing, GRIDs, and Virtualization},
  year = {2025},
  month = {Apr},
  pages = {100--107},
  isbn = {978-1-68558-258-6},
  url = {https://www.thinkmind.org/library/CLOUD_COMPUTING/CLOUD_COMPUTING_2025/cloud_computing_2025_2_60_20036.html}
}

@misc{reuther2024interactiveurgenthpcchallenges,
      title={Interactive and Urgent HPC: Challenges and Opportunities}, 
      author={Albert Reuther and Nick Brown and William Arndt and Johannes Blaschke and Christian Boehme and Antony Chazapis and Bjoern Enders and Robert Henschel and Julian Kunkel and Maxime Martinasso},
      year={2024},
      eprint={2401.14550},
      archivePrefix={arXiv},
      primaryClass={cs.DC},
      url={https://arxiv.org/abs/2401.14550}, 
}

\end{document}